\documentclass[useAMS,usenatbib]{mn2e}
\usepackage{times}
\usepackage{graphics}
\usepackage{graphicx}
\usepackage{color}
\usepackage{verbatim}
\usepackage{amsmath}
\usepackage{amssymb}
\usepackage{mathtools}
\usepackage{subfigure}
\usepackage{float}
\usepackage{tabularx}
\usepackage{supertabular}
\usepackage{rotating}
\usepackage{longtable}
\usepackage{dpfloat, booktabs}
\usepackage{color}
\usepackage{multicol}
\usepackage{graphicx}
\usepackage{multirow}               
\usepackage{amsfonts}

\usepackage{pifont}

\usepackage{wrapfig}
\usepackage{subfig}   

\definecolor{Mygrey}{gray}{0.75}

\title[Molecular Gas in NGC~5033] {Molecular line ratio diagnostics and gas kinematics in the AGN host Seyfert galaxy NGC~5033}
\author[S.\ Topal]  {Sel\c{c}uk Topal$^{1}$\thanks{E-mail:
    selcuktopal@yyu.edu.tr}\\
    $^{1}$Yuzuncu Yil University, Department of Physics, Van, 65080,
  Turkey\\}

\begin{document}
\date{Accepted . Received ; in original form }
\pagerange{\pageref{firstpage}-\pageref{lastpage}} \pubyear{2021}
\maketitle
\label{firstpage}
%
% Abstract
%
\begin{abstract}
Multiple molecular lines are useful for studying the physical properties of multiphase star-forming gas in different types of galaxies. We probe the molecular gas throughout the disc of the spiral galaxy NGC~5033, hosting an active galactic nucleus (AGN), using multiple low-$J$ CO lines [$^{12}$CO(1--0, 2--1, 3--2 and $^{13}$CO(1--0, 2--1)] and dense gas tracers [HCN(1--0) and HCO$^{+}$(1--0)]. Firstly, we determine the ratios of the integrated intensity maps and the ratio of intensities in position velocity diagrams. Secondly, we obtain the ratios of CO lines and high-density tracers at the centre; and thirdly, we model these line ratios using a radiative transfer code. Line ratio diagnostics reveal that the south of the gaseous disc contains cooler gas than the northern part, and the centre hosts warmer and less tenuous gas with a similar dense gas fraction compared to most galaxies of similar type. Our model results mostly agree with the empirical ones in the sense that the central region of NGC~5033 harbours warmer gas than that in the centres of normal spirals and lenticulars without showing AGN activity. Finally, the beam-averaged total molecular gas mass and gas surface density along the galaxy's major axis show a radial gradient, i.e. increasing from the outskirts up to the central region of size $1$~kpc where there is a depression in both gas mass and surface density.

\end{abstract}
%
% Introduction
%
\section{Introduction}
\label{ses:intro}
After realizing that the Milky Way was not the only `island universe' in the cosmos, our understanding of the Universe 
has rapidly expanded as the Universe itself \citep{lem27,hub29,lem31a,lem31b,deva74,tom77,kor79,kor12}. 
There are different types of galaxies with a distinct appearance, luminosity, colour, surface 
brightness, gas content, level of star formation (SF) and activity of their nuclei. `Red and dead' elliptical and lenticular galaxies (i.e. early-types) suffer from a lack of molecular gas, while spirals, like our own Milky Way, are rich in both gas and dust. The central region of galaxies play a crucial role in the dynamics, structure and evolution of galaxies. Now we know that almost all massive galaxies host supermassive black holes (SMBHs) at their centre and energy feedback from central quasar-like events can affect galaxy evolution \citep{kor95}. 

Seyfert galaxies, hosting active galactic nuclei (AGNs), have morphologies similar to spiral galaxies, but with exceptionally 
bright star-like cores. Their nuclei manifest powerful broad emission lines of higher excitation, and they are subdivided into two groups called Seyfert~$1$ and $2$. Studying the ISM of such galaxies through the use of molecular emission lines allows us to better understand SF processes in different types of galaxies. 

Observation and analysis of molecular emission lines enable us to shed light on some outstanding questions regarding SF processes and galaxy evolution. About $200$ molecules have now been detected in the interstellar medium (ISM)\footnote{https://zeus.ph1.uni-koeln.de/cdms/molecules}. Multiple line ratio diagnostics provide us with a great opportunity to study the physics of molecular gas in detail. The upper level of the CO $J=1\rightarrow0$ transition lies $\approx~5.5$~K above the ground state and is relatively easy to detect in molecular clouds where the temperature is typically near $10$~K or higher. CO is also the second most abundant molecule in the ISM after molecular hydrogen (H$_{2}$). CO, therefore, offers an indirect probe for the physics of star-forming gas clouds where H$_{2}$ is hard to observe directly.

%
% General properties of NGC5033
%		
 \begin{table}
 \begin{center}
 \caption{Basic parameters for NGC~5033} %% no full stop at the end of caption
 \begin{tabular}{@{}l l r r r r r@{}}%{lll} 
 \hline  %% rule at top
 %\tablenotemark{a} 
      Property & Value & Ref. \\ \hline
       Type & SA(s)c & a\\
       RA~(J2000) & $13^{\rm h}13^{\rm m}27.5^{\rm s}$ & a\\
       Dec~(J2000) & $36^{\rm d}35^{\rm m}38.2^{\rm s}$ & a\\
       V$_{\odot}$~(km~s$^{-1}$) & $875$ & a\\
       Major diameter (arcmin) & $1.3$ & a\\
      Minor diameter (arcmin) & $0.4$ & a\\
       Distance (Mpc)&$14.7$& b\\
       Position angle (degree) & $171.8$ & b\\
       Inclination (degree) & $66.6$ & b\\	
       Activity class of AGN & S1.8 & b\\	
       SFR~($M_\odot$~yr$^{-1}$) & $0.12$ & c\\ 
 \hline  %% rule at bottom
 \end{tabular}
   \label{tbl:ngc5033}
   \end{center}
    $^{\rm a}$ Nasa/Ipac Extragalactic Database (NED);
  $^{\rm b}$ HyperLEDA \citep{mak14};
  $^{\rm c}$ Max Planck Institute for Astrophysics-Johns Hopkins University Data Release~8 (MPA-JHU DR8);
 \end{table}
 %\footnote{https://leda.univ-lyon1.fr/}

NGC~5033 is a nearby spiral galaxy with Seyfert~$1.8$ type nucleus \citep{ver06} seen at high inclination (Table~\ref{tbl:ngc5033}) with a point-like central X-ray source \citep{tera99}. The galaxy is located at a distance of about $14.7$~Mpc (HyperLEDA; \citealt{mak14}), and is a member of a group \citep{fou92}. General properties of NGC~5033 are listed in Table~\ref{tbl:ngc5033}. We study the physics of the molecular gas in NGC~5033 using multiple low-$J$ CO lines, i.e. $^{12}$CO(1--0), $^{12}$CO(2--1), $^{12}$CO(3--2), $^{13}$CO(1--0), and $^{13}$CO(2--1), and high-density tracers, i.e. HCN(1--0) and HCO$^{+}$(1--0). The number of molecular emission lines available to the current study and the methods used (i.e. empirical line ratio analysis and modelling), make NGC~5033 to an exemplary target to improve our understanding of SF in different types of galaxies, particularly near their centre. We probe physics of the molecular gas across the disc of the galaxy as follows. We first apply empirical line ratio diagnostics along with the non-local thermodynamic equilibrium (non-LTE) code {\sc RADEX} \citep{van07} involving both CO and high density tracers to explore the properties of the molecular gas in the galaxy. Secondly, we study any radial variations in the integrated line intensity, molecular gas mass and surface density throughout the galaxy's major-axis.

Our goals are i) to constrain the physics of 
the central gas and its kinematics over the disc of NGC~5033; ii) to compare our results 
with those of literature studies, particularly those using similar line ratios and the same 
radiative transfer code; iii) discuss possible effects of 
stellar feedback in shaping the ISM at the centre of NGC~5033. 
The sections of this paper are assembled in the following way. The literature data are presented in Section~\ref{sec:lite}. 
The imaging and analysis is given in Section~\ref{sec:analy}. A quantitative analysis of the empirical line 
ratios through the use of {\sc RADEX} is described in Section~\ref{sec:mod}. 
Section~\ref{sec:resdis} presents the main results and further discussions. 
We, finally, conclude briefly in Section~\ref{sec:conc}.

%
% General properties of NGC5033
%		
 \begin{table*}
\begin{center}
 \caption{CO line data} %% no full stop at the end of caption
  \begin{tabular}{@{}l l l c c l c@{}}%{lll} 
% \begin{tabular}{lllllc}
\hline  %% rule at top
      Transition &Rest Freq.& Telescope & Beam size & Linear size$^{a}$ & Reference & Info \\ 
			&(GHz)&&(arcsec)&(kpc)&&\\ \hline 
  			$^{12}$CO(1--0)& $115.3$ & BIMA+$12$m&$6.1\times5.4$&$0.4$& \citealt{hel03}& IM\\\\
			$^{12}$CO(2--1) & $230.5$ &\multirow{2}{*}{JCMT~$15$m}&\multirow{2}{*}{22}&$\multirow{2}{*}{1.6}$&$\multirow{2}{*}{\citealt{pa98}}$&$\multirow{2}{*}{SP}$\\
			$^{13}$CO(2--1) & $220.4$ &&&&&\\ \\
			$^{12}$CO(3--2)& $345.8$ & JCMT~$15$m&$14.5$&$1.0$& \citealt{wil12}&RM\\\\
			$^{12}$CO(3--2) & $345.8$ &HHT~$10$m&$22$&$1.6$& \citealt{mao10}&SP\\	  	
 \hline %% rule at bottom
 \end{tabular}
   \label{tbl:datapro}
 \end{center}
 $^{\rm a}$ Linear size of the beam. Information on the data aquisition method is given in the last column; i.e. IM (interferometric map), RM (raster map), and SP (single dish pointing), respectively.
 \end{table*}

\section{Literature data}
\label{sec:lite}
The literature CO data for NGC~5033 were taken from various sources. Complete maps in $^{12}$CO(1--0) and $^{12}$CO(3--2) were taken from the Berkeley-Illinois-Maryland Association Survey Of Nearby Galaxies survey (BIMA SONG; \citealt{hel03}) and the James Clerk Maxwell Telescope (JCMT) Nearby Galaxies Legacy Survey (NGLS; Project ID MJLSN05; \citealt{wil12}), respectively. The BIMA SONG and JCMT maps in $^{12}$CO(1--0) and $^{12}$CO(3--2) lines cover areas of $5.7$~x~$5.6$ and $5.3$~x~$4.3$~arcmin$^{2}$ over the galaxy, respectively. The $^{12}$CO(1--0) and $^{12}$CO(3--2) data cubes have beam sizes of $6.1$~x~$5.4$ and $14.5$~x~$14.5$~arcsec$^{2}$, respectively. We use these data cubes to create CO integrated intensity maps and position velocity diagrams (PVDs) with the highest resolution possible (Section~\ref{sec:momandpvd}). Additional single-dish spectra of $^{12}$CO(2--1) and $^{13}$CO(2--1) transitions were taken from \citet{pa98} and a $^{12}$CO(3--2) spectrum was taken from the $10$m Heinrich-Hertz Telescope Survey \citep{mao10}. The single-dish spectra of $^{12}$CO(2--1), $^{13}$CO(2--1), and $^{12}$CO(3--2) were obtained with the same beam size of $22$~arcsec. More details on the data described above can be found in the associated papers. Basic information about the literature data is listed in Table~\ref{tbl:datapro}.

\section{Analysis and imaging}
\label{sec:analy} 

\subsection{Integrated line intensities and gas mass}
\label{sec:inten}

The $^{12}$CO(2--1), $^{13}$CO(2--1) and $^{12}$CO(3--2) spectra have a beam size of $22$~arcsec, i.e. the lowest angular resolution in the CO data available for this study, and the reported unit is main beam brightness temperature, $T_{\rm mb}$, in Kelvin (K). However, the interferometric $^{12}$CO(1--0) data cube is in units of Jy and has a beam size of $6.1$~x~$5.4$~arcsec$^{2}$. It is, therefore, necessary to convolve the $^{12}$CO(1--0) cube to the $22$~arcsec beam and make the unit conversion from Jy to $T_{\rm mb}$. We performed a beam convolution to the cube using {\sc MIRIAD} \citep{st95} task $\textit{convol}$. We then estimated the unit conversion factor of $0.19$~K/(Jy/beam) using {\sc MIRIAD} task $imstat$, and applied the conversion factor to the cube. Finally, we extracted the CO(1--0) integrated spectrum from the area of $22$~arcsec~$\times~22$~arcsec at the centre of the galaxy using {\sc MIRIAD} task $\textit{imspect}$.

\begin{figure*}
%\vspace{-50pt}
\begin{center}
  \includegraphics[width=7.0cm,clip=]{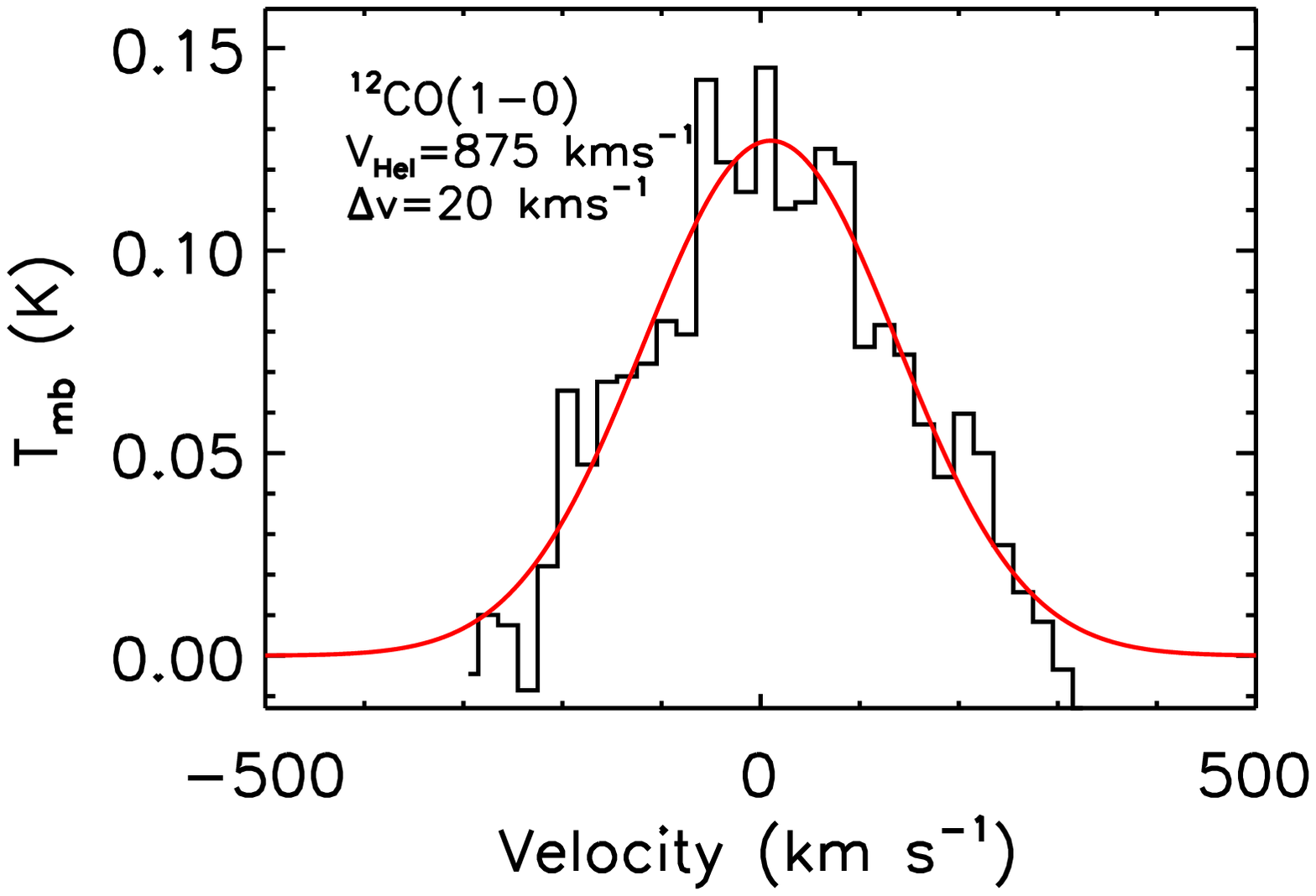}
   \includegraphics[width=7.0cm,clip=]{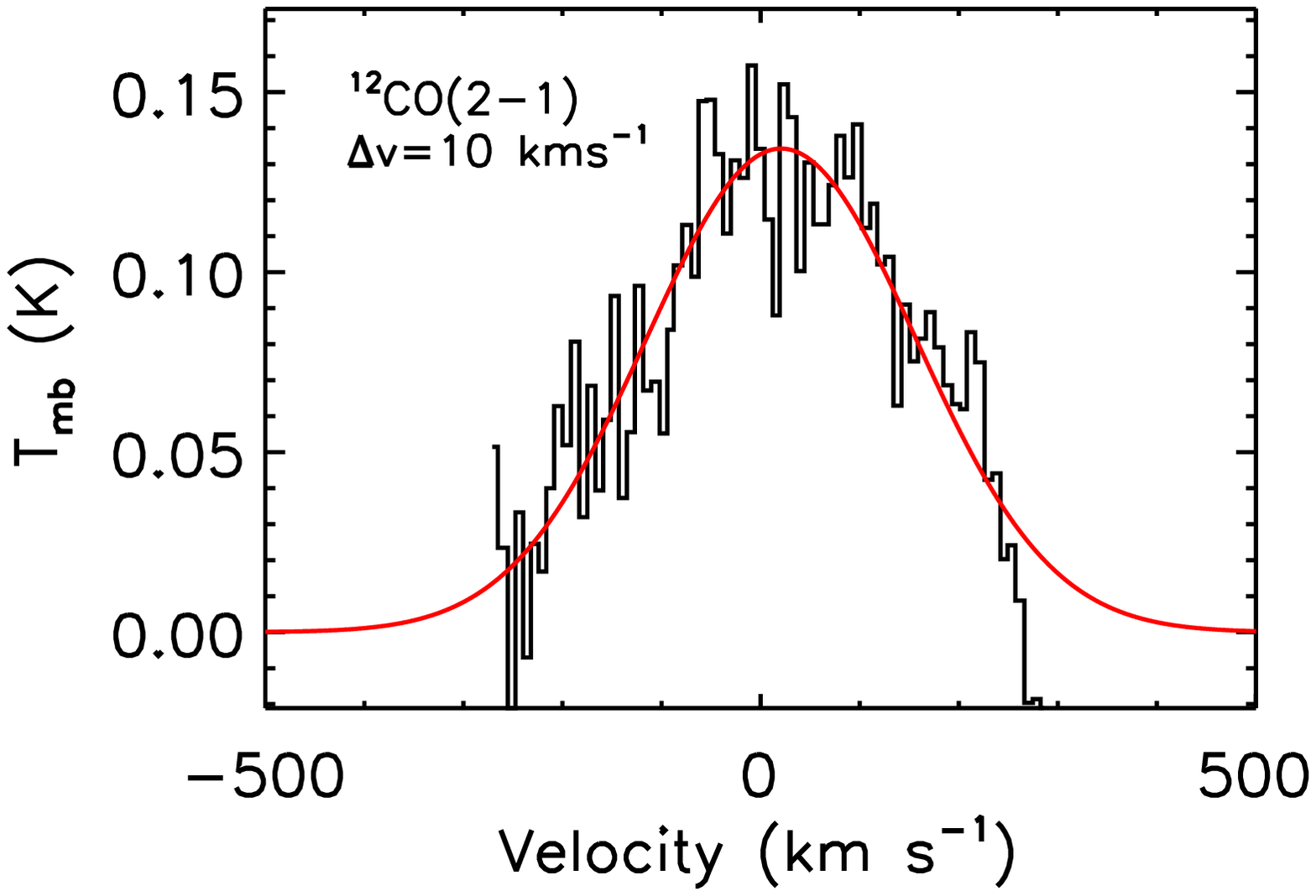}\\
  \includegraphics[width=7.0cm,clip=]{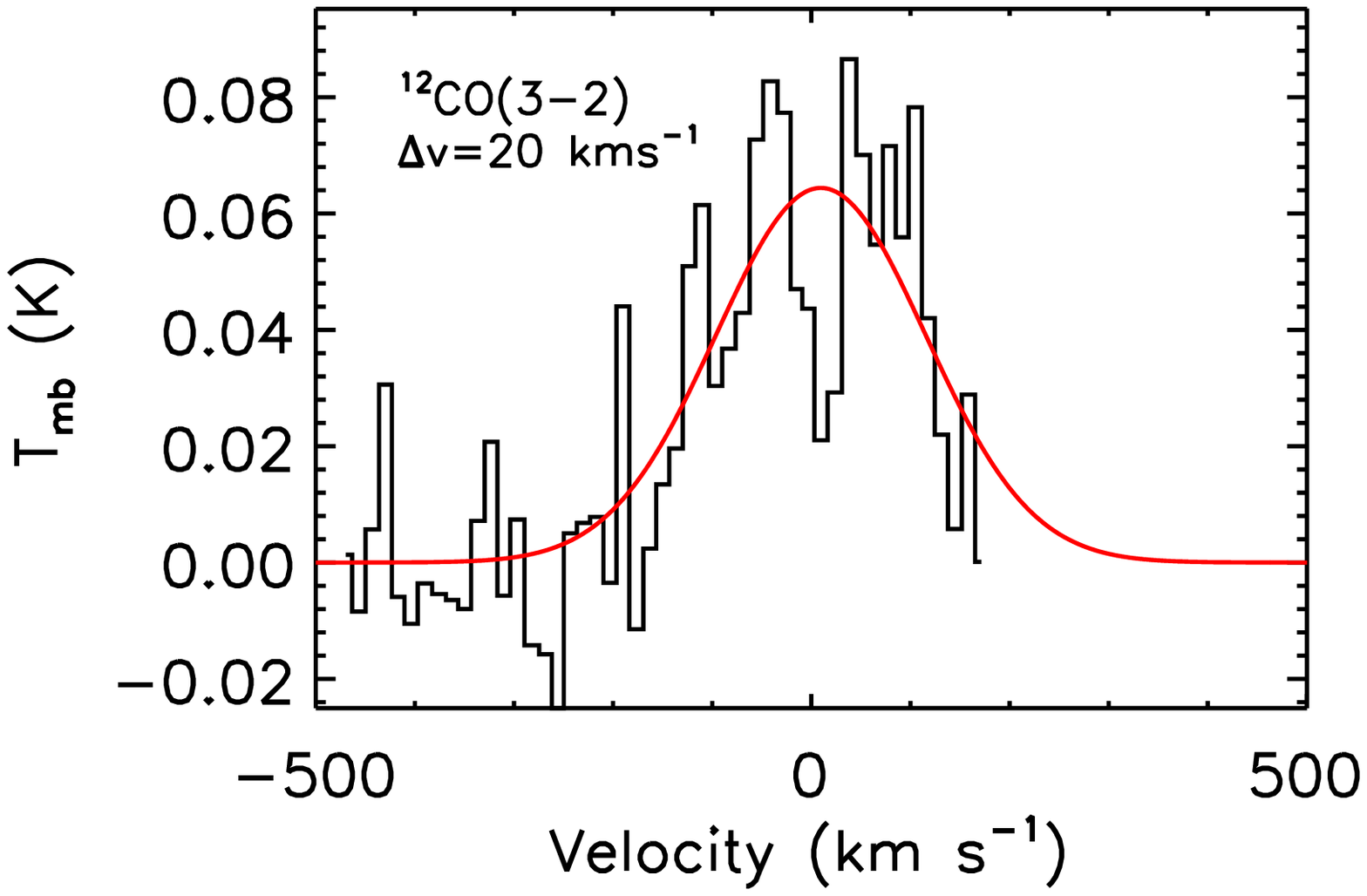}
   \includegraphics[width=7.0cm,clip=]{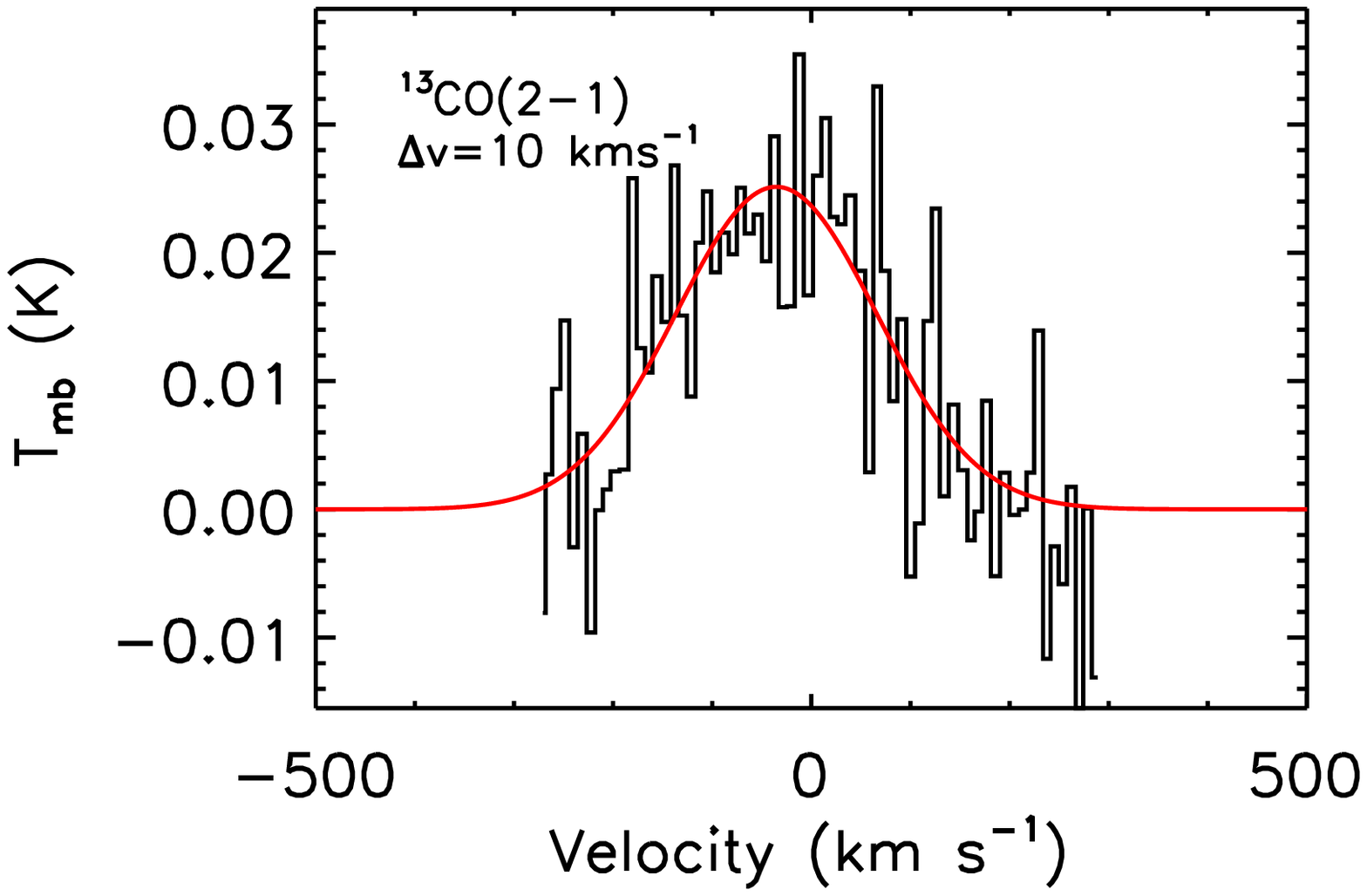}\\
  \caption{Integrated spectra of CO lines for the common beam size of $22$~arcsec. Gaussian fits are overlaid (red). Heliocentric velocity (upper left panel) and the velocity resolution (upper left corners of each panel) are indicated.}
    \label{fig:specfig}
\end{center}
\end{figure*}

All of the $22$~arcsec CO spectra obtained at the centre of the NGC~5033 are shown in Figure~\ref{fig:specfig}. 
Gaussian fits were applied to calculate the integrated line intensities for all spectra. The {\sc MPFIT} package \citep{ma09} was used to optimize the fits. The fitting parameters with the smallest absolute $|1-\chi^2|$ value were considered as the best fit. The intensities calculated for each transition are listed in Table~\ref{tbl:quan}. The beam-averaged total fluxes $F$ (W~m$^{-2}$) per unit area and total intensities $I$ (W~m$^{-2}$~sr$^{-2}$) were calculated using standard expressions \citep{ba04}. The values of total intensities and total fluxes are listed in Table~\ref{tbl:quan}.

% Integrated intensities at the center of NGC5033
%		
 \begin{table*}
\centering
 \small
 \caption{$22$~arcsec beam averaged quantities at the centre of NGC~5033} %% no full stop at the end of caption
 \begin{tabular}{c c c c c c c}
\hline  %% rule at top
 Line & $\int T_{\rm mb}$ d$v$ & Intensity & Flux & FWHM  \\
    &(K~km~s$^{-1}$) & (W~m$^{-2}$~sr$^{-1}$) & (W~m$^{-2}$) & (km~s$^{-1}$)\\
    \hline
      $^{12}$CO(1--0)&$40.8\pm2.7$  &$6.37\,\pm\,0.42\times10^{-11}$ & $8.22\,\pm\,0.54\times10^{-19}$ & $301.6\,\pm\,23.6$\\
      $^{12}$CO(2--1) &$45.8\pm2.7$ &$5.72\,\pm\,0.34\times10^{-10}$ & $7.38\,\pm\,0.43\times10^{-18}$ & $320.5\,\pm\,16.9$\\
      $^{12}$CO(3--2) &$17.1\pm2.7$ &$7.21\,\pm\,1.14\times10^{-10}$ & $9.30\,\pm\,1.47\times10^{-18}$ & $249.5\,\pm\,28.1$\\
      $^{13}$CO(2--1) &$\phantom{0}6.4\pm0.8$&$6.99\,\pm\,0.87\times10^{-11}$&$9.01\,\pm\,1.13\times10^{-19}$&$238.6\,\pm\,29.3$\\	
 \hline %% rule at bottom
 \end{tabular}
 \label{tbl:quan}
 \end{table*}

\begin{figure*}
%\vspace{-50pt}
\begin{center}
\includegraphics[width=6.0cm,clip=]{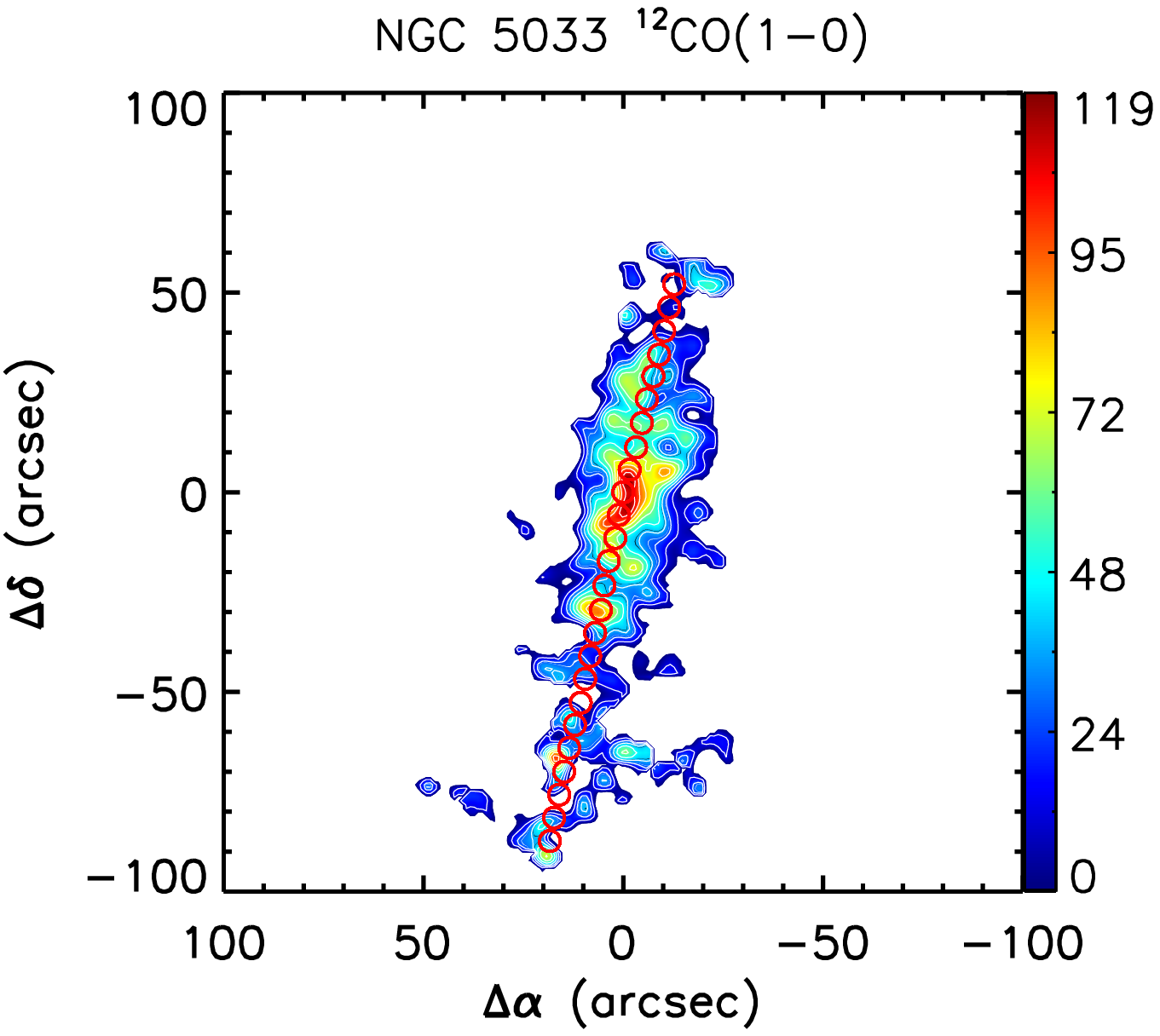}
   \hspace{-40pt}
  \includegraphics[width=6cm,clip=]{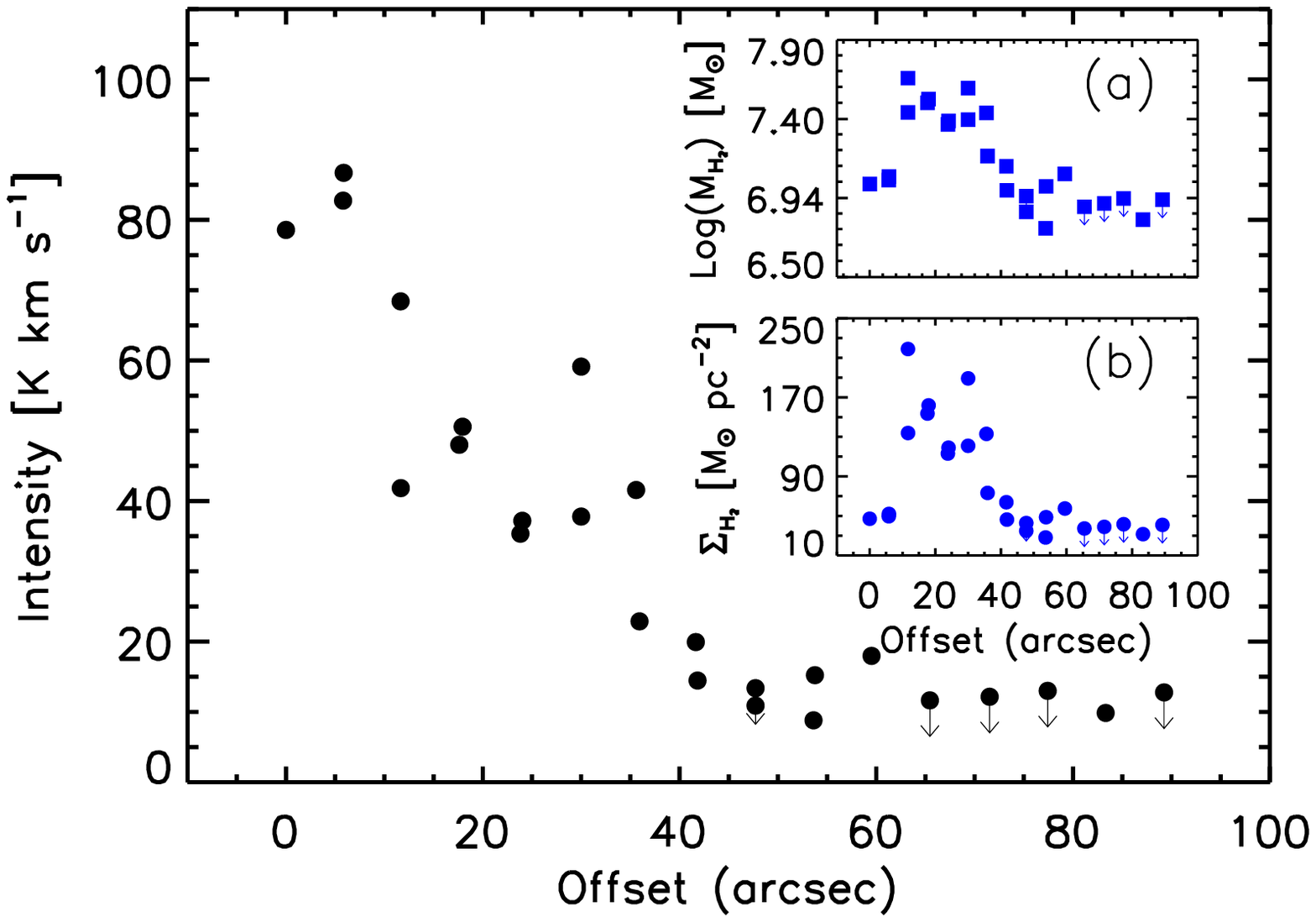}
  \hspace{-40pt}
   \includegraphics[width=6cm,clip=]{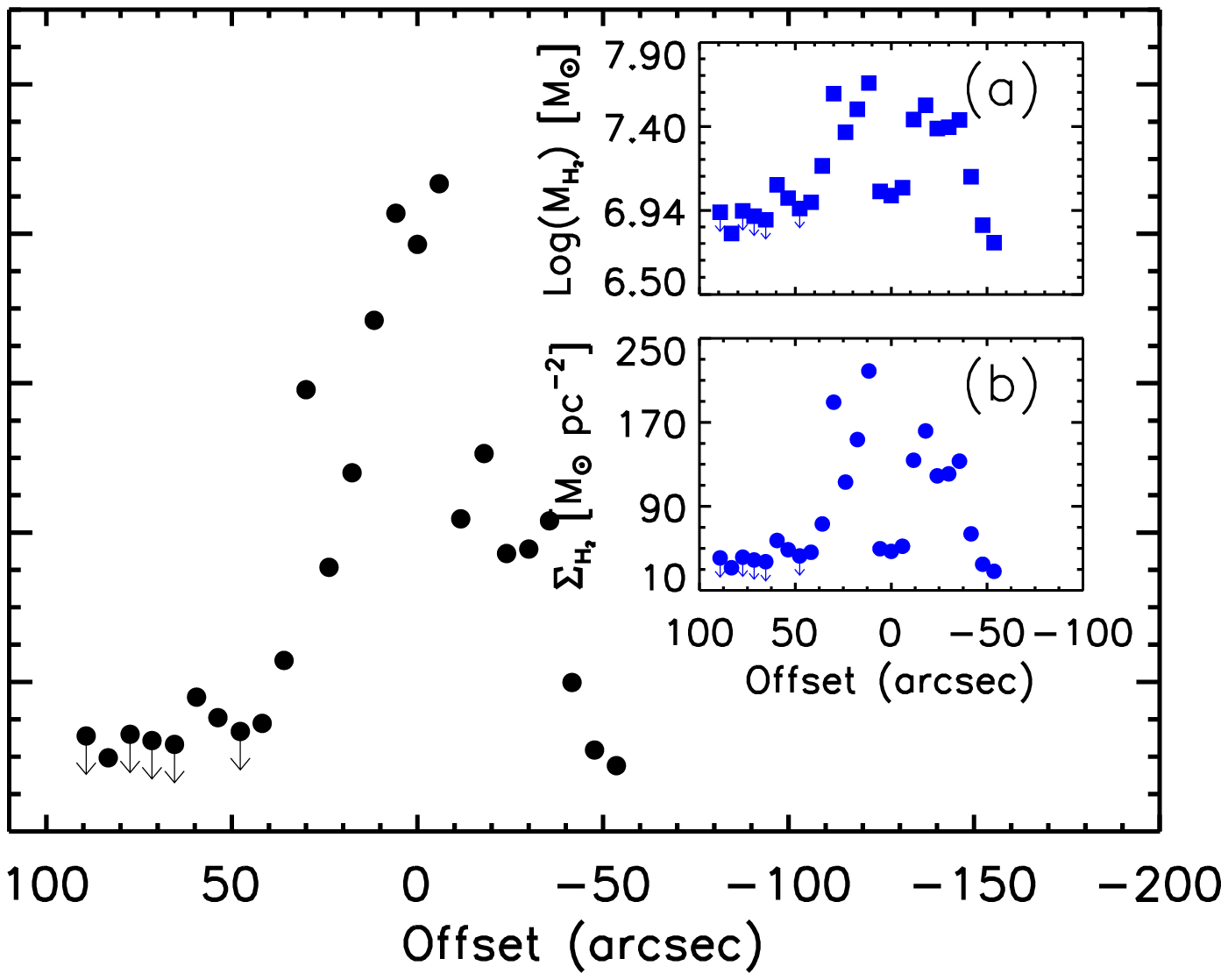}\\
  \caption{Integrated intensity, beam-averaged molecular gas mass, and gas surface density are shown in selected regions across the galaxy's disc. The regions along the major-axis are represented by red circles on the CO(1--0) moment 0 map (left panel). Each red circle has a diameter of $6$~arcsec. The variation of beam-averaged quantities as a function of distance from the galactic centre is shown in the large middle and right panels. Zero offset represents the galaxy centre (Table~\ref{tbl:ngc5033}) in all panels. The black filled circles in the middle and right panels represent the data for the integrated intensity. The blue filled squares in the inserted panels $a$, and the blue filled circles in the inserted panels $b$ represent the gas mass and surface density, respectively. In the right panel and its inserted panels $a$ and $b$, the southern part of the disc is to the left, while the northern part is to the right. The data points with downward arrows show the upper limits of the integrated intensity, i.e. $3\times\sigma_{\rm rms}\times \rm{FWHM}$ (see Section~\ref{sec:inten}).}
    \label{fig:intenmassden}
\end{center}
\end{figure*}

We defined $25$ regions side by side over the major-axis so that we can probe any changes in the physical properties of the gas as a function of radius without selection bias (see Figure~\ref{fig:intenmassden}). Each circular region has a diameter of $6$~arcsec, i.e. about the beam size of the $^{12}$CO(1--0) data, representing a circular region with a diameter of $400$~pc along the major axis of the galaxy. We first extracted the $^{12}$CO(1--0) spectra in the selected regions and calculated the integrated intensities, as explained above. $^{12}$CO(1--0) was not detected ($\int T_{\rm mb}$ d$v$ $< 3\sigma$, where $\sigma$ is the uncertainty in the integrated intensity) at $5$ positions in the SE of the galaxy's major axis (see Figure~\ref{fig:intenmassden}). For those $5$ positions, we estimated the upper limit to the integrated intensity as $3\times\sigma_{\rm rms}\times \rm{FWHM}$, where $\sigma_{\rm rms}$ is the noise in the spectrum at that position and FWHM is the average Full Width at Half Maximum line width of the $^{12}$CO(1--0) spectra in the detected positions, i.e. $40$~km~s$^{-1}$. The upper limit could also be calculated as $3\times\sigma_{\rm rms}\times \delta v\times\sqrt{N}$, where $\delta v$ is the channel width and $N$ is the number of channels across the line. However, since BIMA $^{12}$CO(1--0) data cube has a channel width of $20$~km~s$^{-1}$ and the average number of channels across the line is about $4$ for the $^{12}$CO(1--0) line profiles at the detected positions, the latter method would give almost identical values to those estimated from the first method.

We estimated the gas mass in units of solar mass ($M_{\odot}$) in each selected region using CO-to-H$_{2}$ conversion factor ($X_{\rm CO}$), the distance to the galaxy, and the measured $^{12}$CO(1--0) integrated red intensities. $X_{\rm CO}$\,=\,$2\times10^{20}$ cm$^{-2}$~(K~km~s$^{-1}$)$^{-1}$ was taken because it is a good choice for the disc of our Galaxy \citep{dame01,abdo10,bol13}. Studies of nearby `normal galaxies' indicate similar values \citep{ros03, bo08, don12}. The discs of Seyfert galaxies show conversion factors, which are within a factor of two consistent with the Galactic disc value \citep{don13}. However, although it is not universal, many studies showed that $X_{\rm CO}$ could be lower by a factor of $5$ to $10$ in the central region of galaxies regardless of galaxy type \citep[e.g.][]{sand13,bol13}. In the central region of the Seyfert galaxy NGC~1068, for example, the conversion factor was found to be $X_{\rm CO} = 0.2-0.4\times10^{20}$ cm$^{-2}$~(K~km~s$^{-1}$)$^{-1}$ \citep{papa99}. However, please note that the value of the conversion factor depends on the method used (i.e. virial mass, dust column density, spectral line modelling), resolution and metallicity (\citealt{bol13} and references therein). Such uncertainties could cause error bars in the estimated nuclear value of $X_{\rm CO}$. 

For NGC~5033 we, therefore, adopted values of $X_{\rm CO}=0.4\times10^{20}$ cm$^{-2}$~(K~km~s$^{-1}$)$^{-1}$ and $X_{\rm CO}=2\times10^{20}$ cm$^{-2}$~(K~km~s$^{-1}$)$^{-1}$ for the central $1$~kpc (i.e. the central three red $6$ arcsec sized regions, see Fig.~\ref{fig:intenmassden}) and for the arms, respectively, where the corresponding values of $\alpha_{\rm CO}\,\equiv\,\Sigma_{\rm H_{2}}$ / $I_{\rm 1-0}$ are $0.6~M_{\odot}$(K~km~s$^{-1}$~pc$^{2}$)$^{-1}$ and $3.2~M_{\odot}$(K~km~s$^{-1}$~pc$^{2}$)$^{-1}$, respectively \citep{nar12}. The expression used to estimate the molecular gas mass is shown below.
\begin{equation}
  \frac{M_{{\rm H}_2}}{M_\odot}=C\,\bigg(\frac{I_{\rm 1-0}}{{\rm K}\,{\rm km}\,{\rm s}^{-1}}\bigg)\,,
  \label{eq:mass}
\end{equation}
where $I_{1-0}$ is the main beam brightness $^{12}$CO(1--0) integrated line intensity in units of [K~km~s$^{-1}$], and $C = 1.3\times10^5$ for $X_{\rm CO} = 0.4\times10^{20}$ cm$^{-2}$~(K~km~s$^{-1}$)$^{-1}$ and $C = 6.6\times10^5$ for $X_{\rm CO} = 2\times10^{20}$ cm$^{-2}$~(K~km~s$^{-1}$)$^{-1}$. The gas surface density, $\Sigma_{\rm H_{2}}$, was also estimated for each selected region using $I_{\rm 1-0}$ and $X_{\rm CO}$. The integrated $^{12}$CO(1--0) intensities, and the beam-averaged $M_{{\rm H}_2}$ and $\Sigma_{\rm H_{2}}$ obtained in the selected regions are shown in Figure~\ref{fig:intenmassden}.

On a different note, any two elements of an interferometer cannot be closer than some minimum distance, causing the problem known as `missing flux' in the data. However, BIMA SONG data cubes for $24$ galaxies (including NGC~5033) were created by incorporating the total flux measurement from the NRAO $12$m telescope to achieve the flux recovery (see \citealt{hel03}). It is therefore important to note that the total mass estimation in the current study is not significantly affected by missing flux.

\subsection{Moment maps and position velocity diagrams (PVDs)}
\label{sec:momandpvd}
\begin{figure*}
\begin{center}
%\vspace{-50pt}
  \includegraphics[width=8.5cm,clip=]{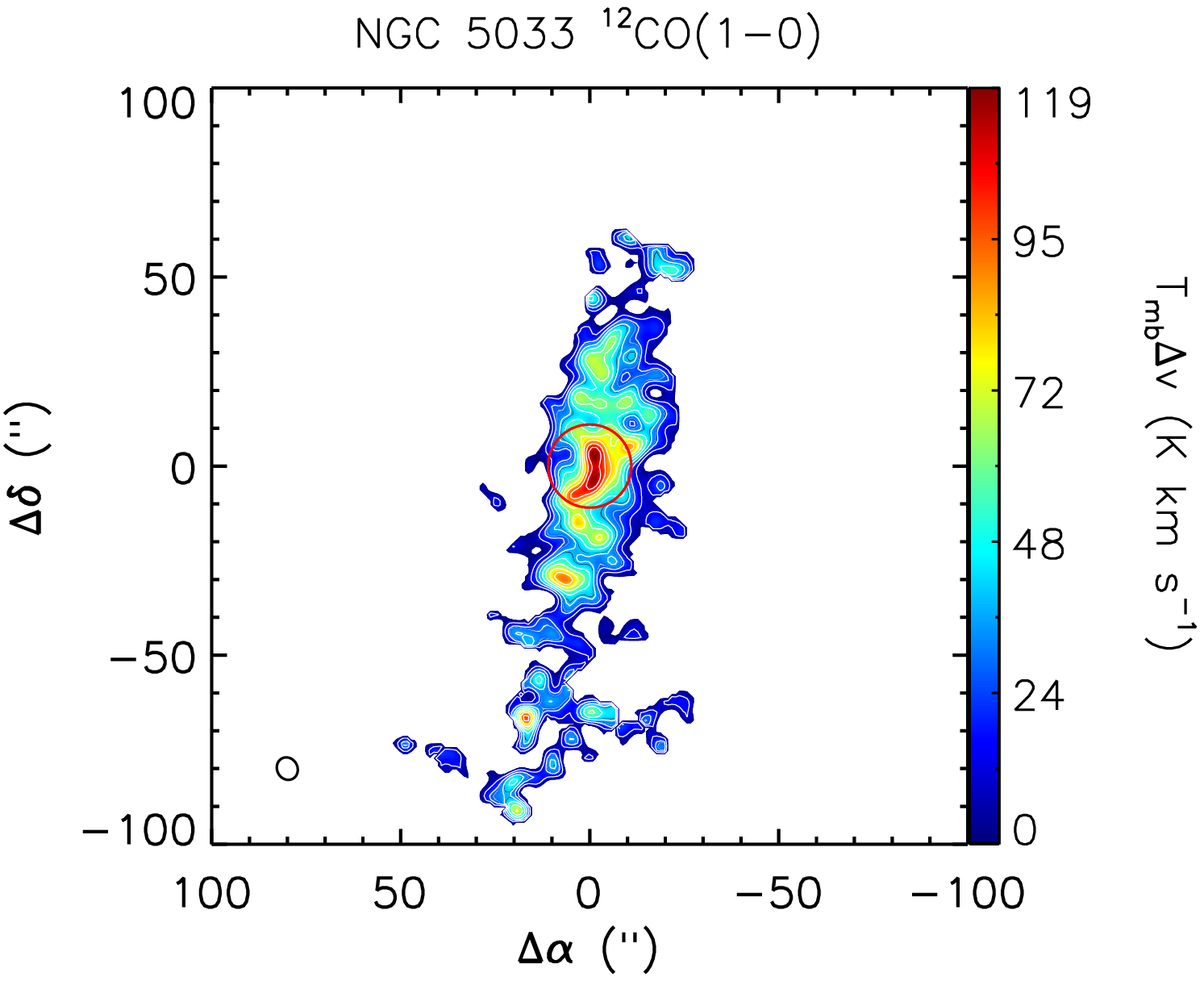}
   \includegraphics[width=8.5cm,clip=]{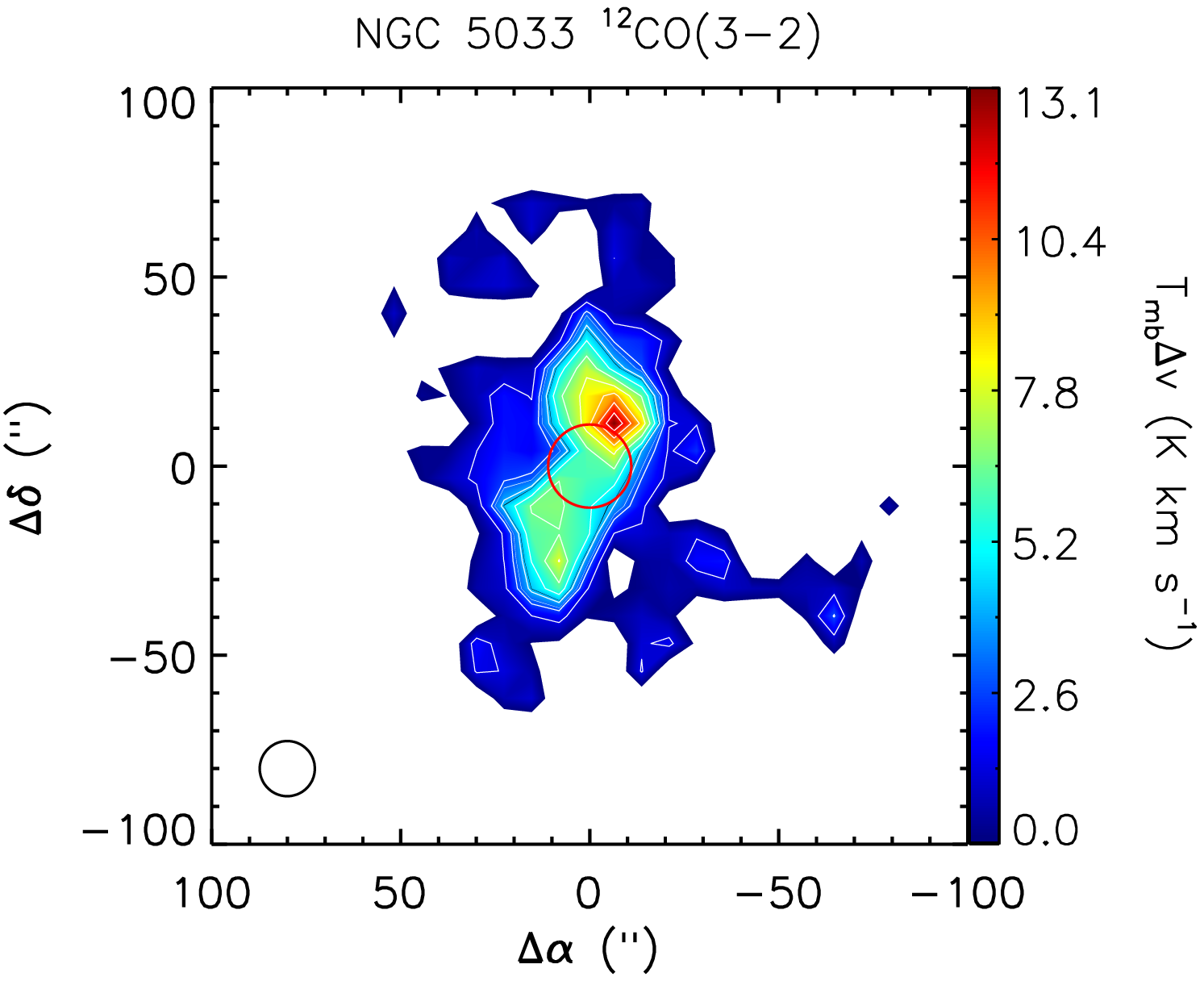}\\
  \includegraphics[width=8.5cm,clip=]{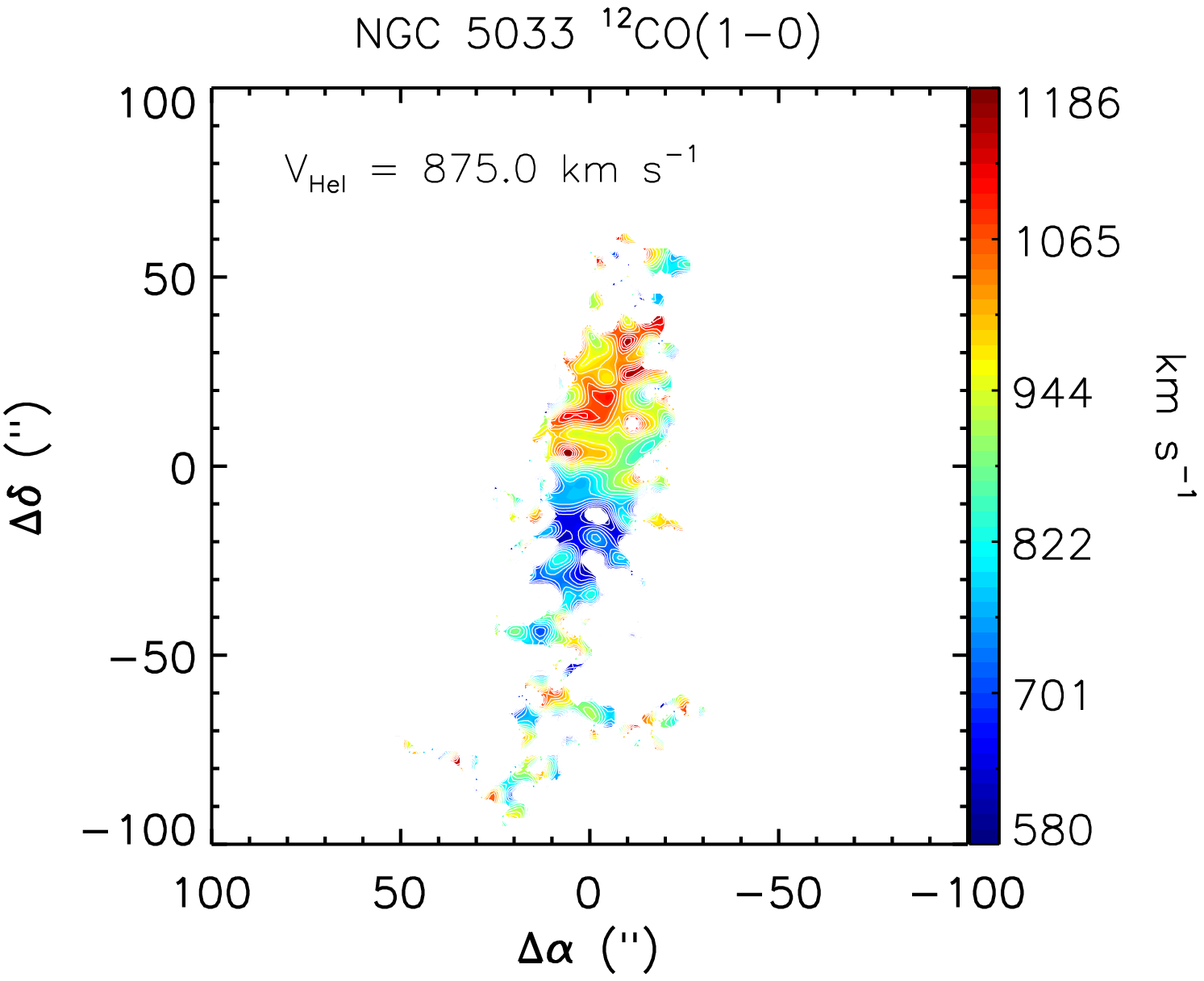}
   \includegraphics[width=8.5cm,clip=]{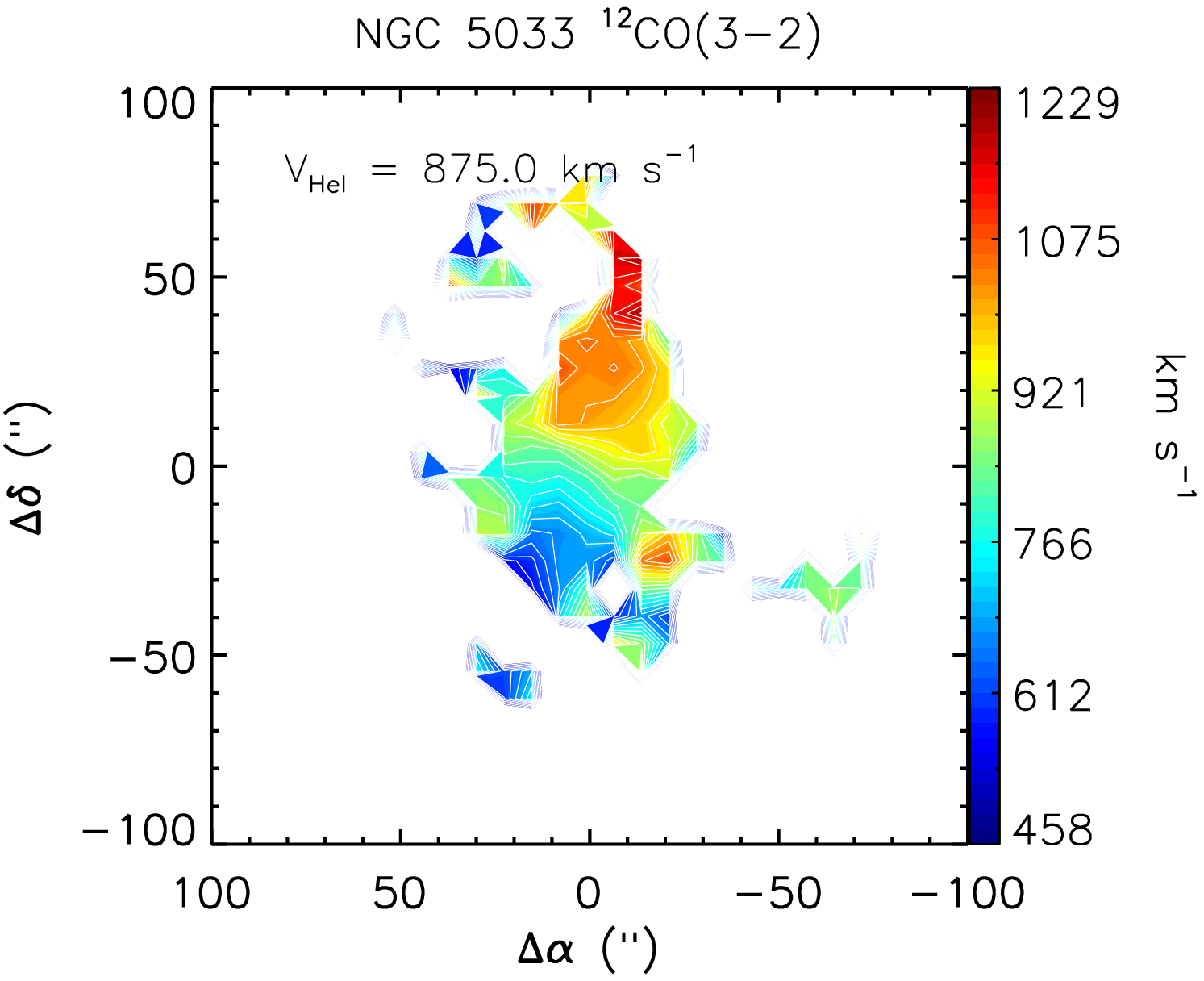}\\
    \includegraphics[width=8.5cm,clip=]{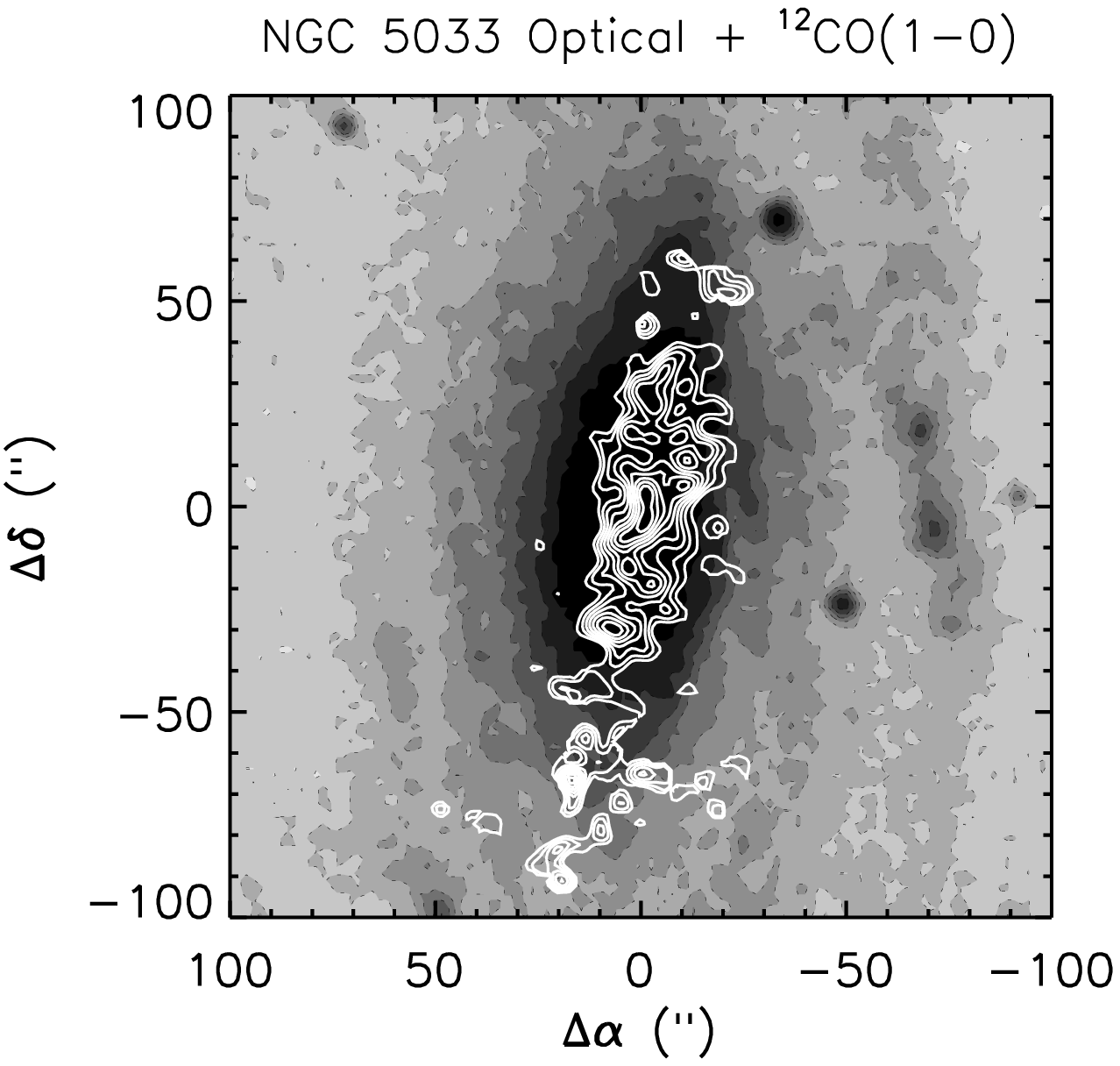}
   \includegraphics[width=8.5cm,clip=]{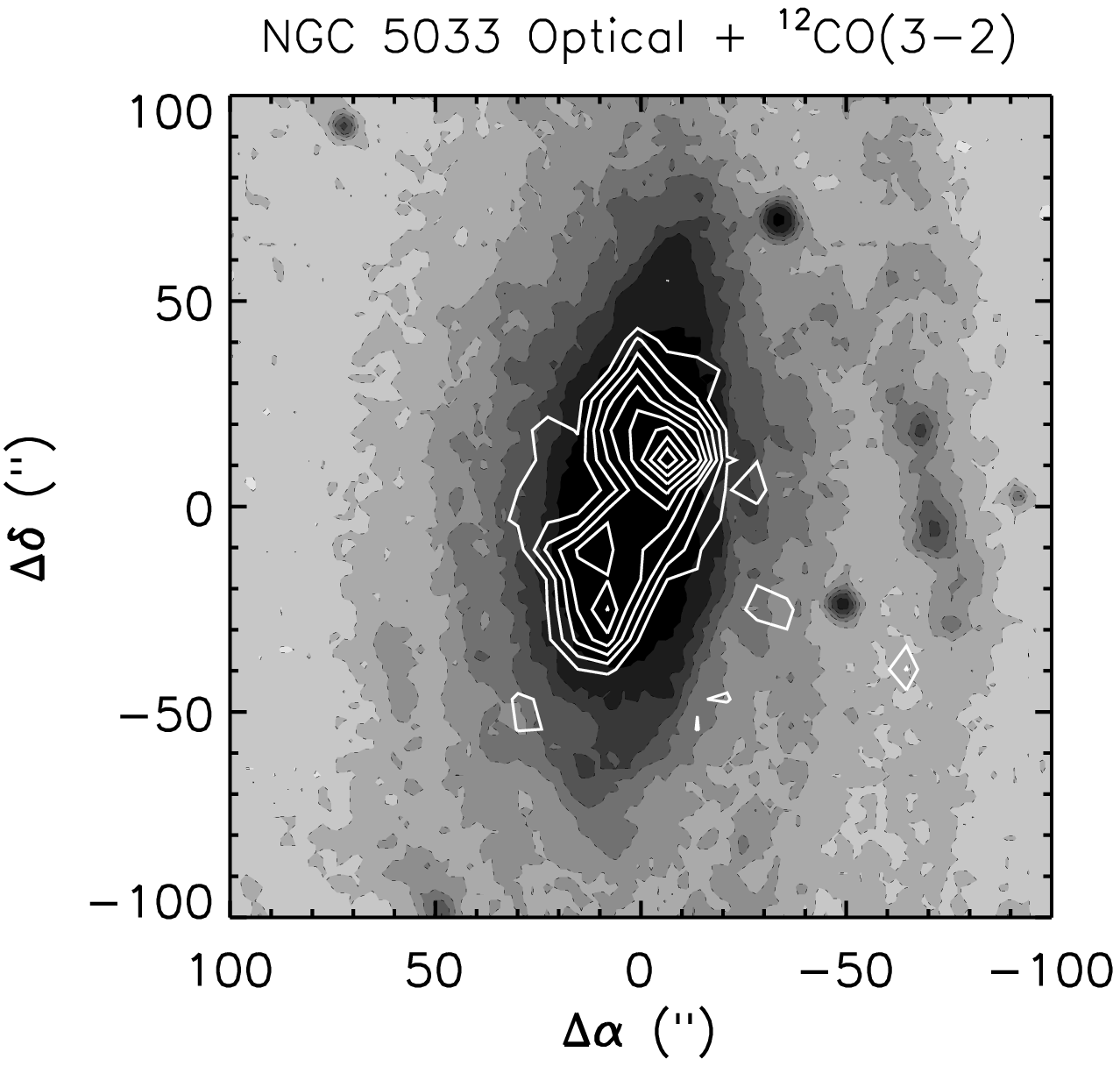}\\
  \caption{{\bf Top}: Moment 0 contour maps for $^{12}$CO(1--0) (left) and $^{12}$CO(3--2) (right), respectively. The black circle in the bottom left of each panel represents the beam sizes. Levels of integrated intensity contours on the moment 0 maps range from $10$ to $100$ per cent of the peak intensity with an intercept of $10$ per cent. The peak intensities in the moment 0 maps of $^{12}$CO(1--0) and $^{12}$CO(3--2) are $119.2$~K~km~s$^{-1}$ and $13.3$~K~km~s$^{-1}$, respectively. The red circles centred on both images have a diameter of $22$~arcsec (i.e. lowest resolution in the adopted CO data set). {\bf Middle}: Moment1 contour maps for $^{12}$CO(1--0) and $^{12}$CO(3--2). Velocity contours on the moment 1 maps are separated by $30$~km~s$^{-1}$. The systemic velocity of $875$~km~s$^{-1}$ is shown in the top left of each panel. {\bf Bottom}: The moment 0 maps (white contours) are superimposed on an optical image of the galaxy (greyscale) taken from the Sloan Digital Sky Survey (SDSS). East is to the left and north is up in all panels.}
  \label{fig:moments}
  \end{center}
\end{figure*}

We obtained integrated line intensities (hereafter moment 0) and velocity maps (hereafter moment 1) by defining a region of contiguous emission in the $^{12}$CO(1--0) and $^{12}$CO(3--2) data cubes (see \citet{topal16} for more details on the procedure). Moment maps of the galaxy are shown in Figure~\ref{fig:moments}.
%

%
% Figure: Moment0 and Moment1 maps 
%

\begin{figure*}
\begin{center}
  \includegraphics[width=8cm,clip=]{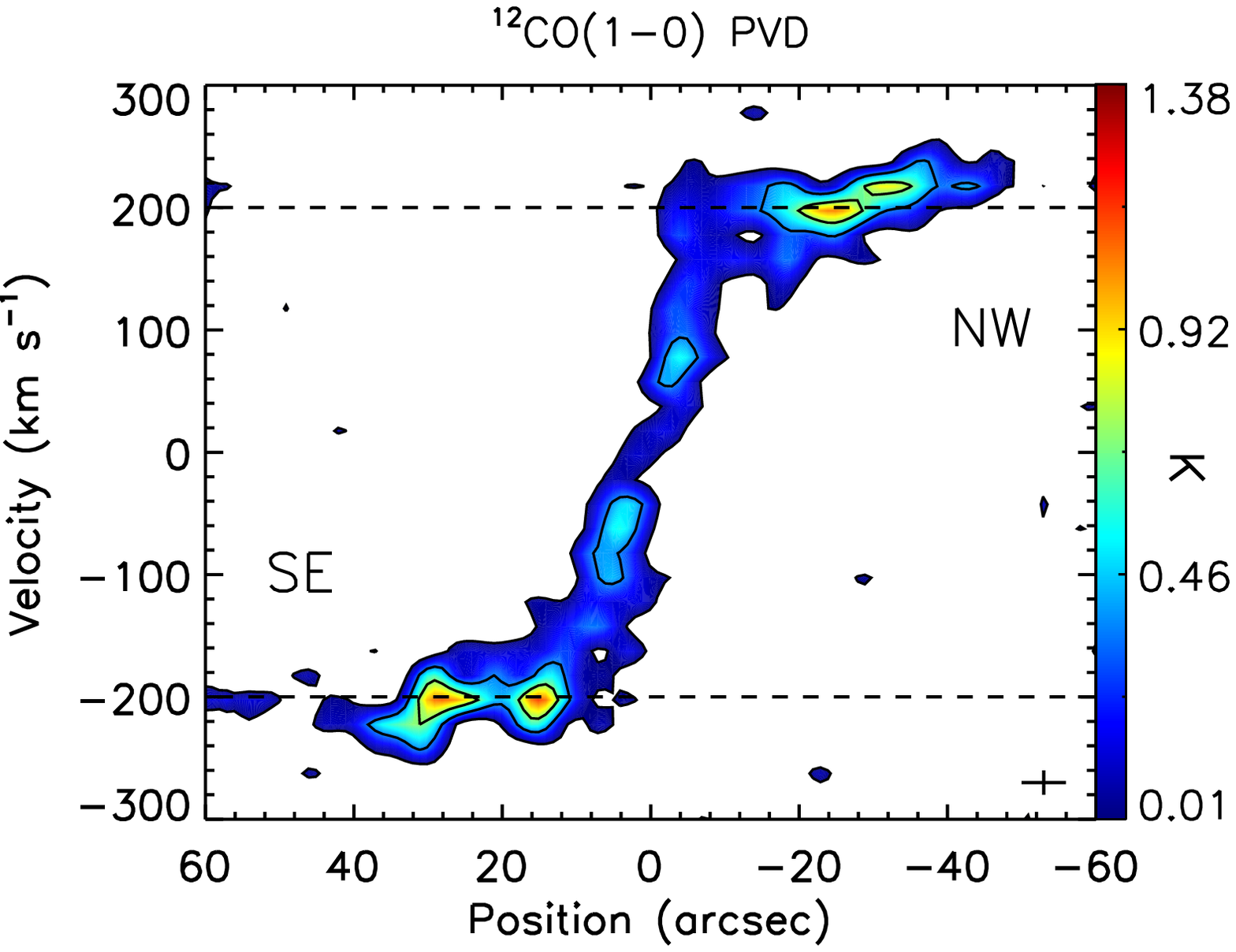}
  \includegraphics[width=8cm,clip=]{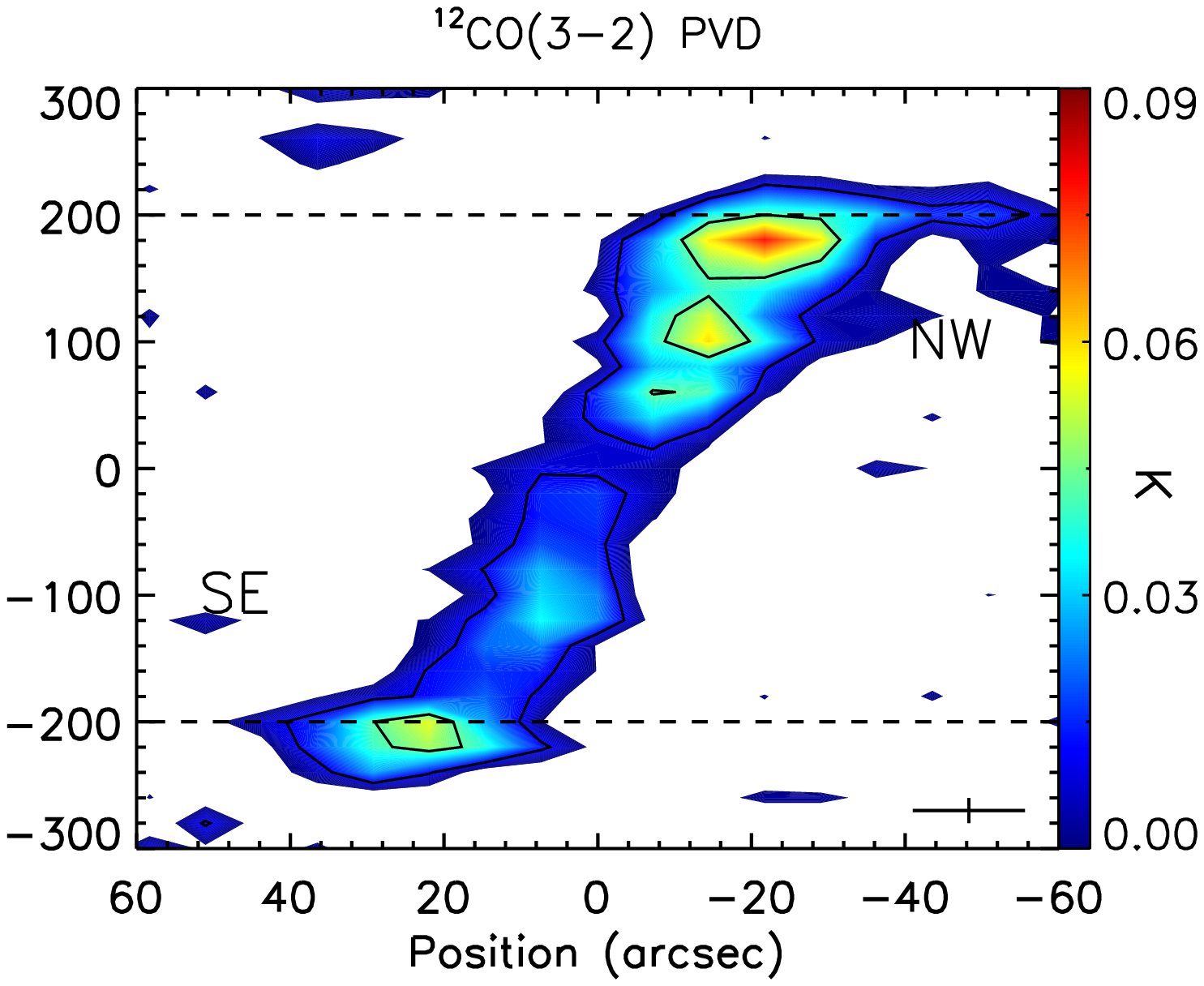} \\
  \caption{Position velocity diagrams (PVDs) of $^{12}$CO(1--0) (left) and $^{12}$CO(3--2) (right). The rms noise (i.e. $1\sigma$) for $^{12}$CO(1--0) is $0.11$~K while the $^{12}$CO(3--2) PVD has an rms noise of $0.01$~K. Black lines on the PVDs represent the contour levels spaced by $3\sigma$ and starting at $2\sigma$. Colour on the contour maps starts at $2\sigma$ and $1\sigma$ for $^{12}$CO(1--0) and $^{12}$CO(3--2), respectively. The rotational velocity of the gas at $200$~km~s$^{-1}$ is indicated with black horizontal dashed lines. The horizontal and vertical lines in the lower right part of each panel indicate the spatial resolution over the major-axis and velocity resolution, respectively.}
  \label{fig:pvds}
  \end{center}
\end{figure*}

We created position velocity diagrams (PVDs) by taking a slice along the major-axis of the galaxy in the fully-calibrated and cleaned data cubes of $^{12}$CO(1--0) and $^{12}$CO(3--2) (see Figure~\ref{fig:intenmassden}, left panel). The $^{12}$CO(1--0) and $^{12}$CO(3--2) data cubes have pixel sizes of $1$ and $7.3$~arcsec, respectively. We first rotated the cubes using the {\sc MIRIAD} task $regrid$ to adjust the pixels to a horizontal grid. We then took a slice along the major-axis of the galaxy, averaging a total of five (for the $^{12}$CO(1--0) cube) and three (for the $^{12}$CO(3--2) cube) pixels in perpendicular direction (i.e. along the minor-axis). The spatial resolution of each PVD was calculated using the beam major-axis and minor-axis, the beam position angle, and the gas position angle (see \citealt{dav13}). The PVDs are shown in Figure~\ref{fig:pvds}.

%
% Figure: PVDs
%

\subsection{Line ratios at the centre and over the disc}
\label{sec:ratmompvd}
We first measured the ratios of the integrated line intensities at the centre. Additional integrated line intensity ratios of $^{12}$CO(1--0) / $^{13}$CO(1--0), HCN(1--0) / $^{12}$CO(1--0) and HCN(1--0) / HCO$^{+}$(1--0) were taken from \citet{kri10}. All in all, the ratios we obtained at the centre of NGC~5033 are $^{12}$CO(2--1) / $^{12}$CO(1--0) (hereafter $R_{\rm 21}$), $^{12}$CO(1--0) / $^{13}$CO(1--0) (hereafter $R_{\rm 11}$), $^{12}$CO(3--2) / $^{12}$CO(1--0) (hereafter $R_{\rm 31}$), $^{12}$CO(2--1) / $^{13}$CO(2--1) (hereafter $R_{\rm 22}$), HCN(1--0) / $^{12}$CO(1--0) (hereafter $R_{\rm DC}$) and finally HCN(1--0) / HCO$^{+}$(1--0) (hereafter $R_{\rm DD}$). The ratios are listed in Table~\ref{tbl:ratiomod}. We then compared the ratios found at the centre of the galaxy to those found in the centre and in the arms of other types of galaxies from the literature (see Figure~\ref{fig:linerat} and Section~\ref{sec:ratio}).

As our second approach to the line ratio analysis, we measured the ratio of the moment 0 maps enabling us to probe the change in the $R_{\rm 31}$ ratios across the disc of the galaxy. We then measured the ratio of the PVD intensities (similar to \citealt{topal16}) allowing us to probe the physical conditions throughout the gas kinematic components of the galaxy, i.e. nuclear disc and/or inner ring (see Section~\ref{sec:mompvd}). The method we used to obtain the ratio moment 0 maps and ratio PVDs is explained below.

Before taking the ratio of the moment maps, we first convolved the $^{12}$CO(1--0) moment 0 map to the beam size of $14.5$~arcsec, i.e. the beam size of the $^{12}$CO(3--2) data. We also applied additional corrections so that both moment 0 maps have the same pixel size and both cover the same area in the sky. The ratio moment map is shown in Figure~\ref{fig:mompvd}. For the PVD intensity ratios we also made sure that the starting and ending velocities, velocity resolution and the area in the sky were the same for both cubes. We then created the PVDs from those identical cubes, by following the same steps explained in Section~\ref{sec:momandpvd}. We finally obtained the ratio of the PVDs (see Figure~\ref{fig:mompvd}).

\section{Modelling}
\label{sec:mod} 
We probe the physics of the molecular gas quantitatively by performing RADEX calculations. RADEX provides a Large Velocity Gradient (LVG) model and creates a parameter space for a user-defined set of physical parameters. We define the model parameters in two ways with a $\chi^2$ and a likelihood method. Further details on the radiative transfer code {\sc RADEX} and the best model identification can be found below.

\subsection{LVG code: {\sc RADEX}}
\label{sec:lvg}

The input parameters for {\sc RADEX} are the gas kinetic temperature $T_{\rm K}$, hydrogen gas volume number density $n$(H$_2$), the carbon monoxide column number density $N$(CO), the line width of the spectrum $\Delta v$, and the intrinsic abundance ratio, [$^{12}$C] / [$^{13}$C]. The radiation of the cosmic microwave background ($T_{\rm CMB}$ = $2.73$~K) is also implemented. We considered line widths of $300$~km/s and $250$~km/s for $^{12}$CO and $^{13}$CO, respectively. Note that the line widths minimally affect the {\sc RADEX} results \citep{van07}. The model grids were created as follows. Given the low-$J$ CO lines available to this study, $T_{\rm K}$ ranges from $5$ to $20$~K with an increment of $1$~K. By doing that, we aim to sample the lower temperature regime better. Beyond $20$~K, $T_{\rm K}$ goes up to $250$~K with an increase of $5$~K in each step. $n$(H$_2$) and $N$(CO) range from $10^{2}$ to $10^{7}$~cm$^{-3}$, and from $10^{13}$ to $10^{21}$~cm$^{-2}$, respectively, in steps of $0.25$~dex. 

In our analysis, we assume that isotopic ($^{12}$C / $^{13}$C) and isotopologue ($^{12}$CO / $^{13}$CO) abundance ratios 
are not different. The [$^{12}$C]/[$^{13}$C] abundance ratio can change dramatically from one galaxy to another, and even in 
different locations over the same galaxy. The abundance ratio may also show a radial gradient, i.e. the ratio increases with the distance from the Galactic centre \citep[e.g.][]{wil92,mil05, yan19}. In our Galaxy it ranges from $20$-$25$ near the centre \citep[e.g.][]{wr94, mul16, cor18} to $\approx70$ near the solar circle \citep[e.g.][]{wil92, mil05}. Additionally, in the far-outer Galaxy the abundance ratio reaches $100$ with a larger error bar and there is an inconsistency in the results obtained from the $J = 1-0$ and $J = 2-1$ transitions \citep{wou96}. An abundance ratio of $50$ was found in Magellanic Clouds \citep{w09}. In starburst nuclei of nearby galaxies NGC~253  and M82 the abundance ratio is [$^{12}$C]/[$^{13}$C] = $40-50$ \citep{hm93b}, while more recently it has been suggested to be even higher, i.e. the lower limits for the abundance ratio are [$^{12}$C]/[$^{13}$C] = $81$ and [$^{12}$C]/[$^{13}$C] = $138$ for NGC~253 and M82, respectively \citep{mar10}. In ultra-luminous infrared galaxies the ratio was found to be $90$ \citep{sd17}. All of these extragalactic values are higher than the value in the centre of the Milky Way. As better representatives to NGC~5033, in the centres of red the Seyfert galaxies NGC~1068 and NGC~4945, the abundance ratios were found to be [$^{12}$C]/[$^{13}$C] = $24-62$ (with an average of $\approx38$) and [$^{12}$C]/[$^{13}$C] = $6-44$ (with an average of $\approx17$), respectively \citep{tang19}. In our model calculations, instead of taking one single absolute value, we therefore considered a range for the abundance ratio, i.e. from $20$ to $90$ in steps of $10$, covering almost the entire range of the abundance ratios reported in the literature.

\subsection{Best model identification}
\label{sec:bestlvg}
We applied  $\chi^2$ and likelihood methods to identify the best model results. $\chi^2$ is defined as,

\begin{equation}
  \chi^{2}\equiv\sum\limits_{i}\bigg(\frac{R_{i,{\rm
      mod}}-R_{i,{\rm obs}}}{\Delta R_{i,{\rm obs}}}\bigg)^2\, 
  \label{eq:chi2}
\end{equation}
where $R_{\rm mod}$ and $R_{\rm obs}$ represent modelled and observed line ratios, respectively, while $\Delta R_{\rm obs}$ 
represents the uncertainty in the observed line ratios. The best model was defined as the model with the smallest reduced $\chi^2$, ($\chi_{\rm r}^2$ = $\chi^2$ / $DOF$, where $DOF$ represents corresponding degrees of freedom, i.e. one less than the number of line ratios). The contour map of $\Delta\chi_{\rm r}^{2}$ = $\chi_{\rm r}^2$ - $\chi_{\rm r, min}^2$ is shown in Figure~\ref{fig:chilike}. 

We created the probability distribution functions (PDFs) of each model parameter (namely $T_{\rm K}$, $n$(H$_2$) and $N$(CO)) by following the procedure described in \citet{kav07}. We first estimated the likelihood of each model parameter by calculating the sum of the $e^{-\Delta\chi^{2} / 2}$ (i.e. the likelihood function) for all possible values of the other two parameters marginalized over the other two. Assuming a uniform prior in the model parameters and Gaussian errors lead prob($\emph{D}~|~\emph{X}$) $\propto$ $e^{-\Delta\chi^{2} / 2}$, where prob($\emph{D}~|~\emph{X}$) is the probability of the data given the model, $\emph{X}$ denotes the parameters in the model and the $\chi^{2}$ is defined in the standard way shown above. Please see Section~2 in \citet{kav07} for more details on the procedure. The PDFs, including the median with the $68$~per cent ($1\sigma$) confidence levels around it, and the most likely model (i.e. the peak) are shown in Figure~\ref{fig:chilike}.

\section{Results and discussion}
\label{sec:resdis}

\subsection{Intensities, gas mass and surface density}
\label{sec:massden}
From the outskirts to the central $1$~kpc, both gas mass and surface density increase. Although the integrated CO line intensity is the highest in the central $1$~kpc (see Fig.~\ref{fig:intenmassden}), the inverse relation between CO integrated line intensity and $X_{\rm CO}$ and the depression in $X_{\rm CO}$ and $\alpha_{\rm CO}$ assumed for the central $1$~kpc region (see Section~\ref{sec:inten}) cause a decrease in gas mass and molecular surface density in the central region of the galaxy (see Figure~\ref{fig:intenmassden}). Given the patchy distribution of the gas beyond an angular galactocentric distance of $\approx50$~arcsec in the south of the galaxy, this causes $^{12}$CO(1--0) emission not being detected in five regions at a $3\sigma$ level, indicated by arrows in Figure~\ref{fig:intenmassden}. However, in the north of the disc, the true extent of the emission (see Section~\ref{sec:momandpvd}) does not reach beyond $\approx50$~arcsec. This indicates that between $50$ to $100$~arcsec from the centre the south of the galaxy tends to a have higher surface density of molecular gas.

\subsection{Moment maps and PVDs}
\label{sec:mompvd}
$^{12}$CO(1--0) is more extended over the galaxy's major axis compared to $^{12}$CO(3--2) (see Figure~\ref{fig:moments}). The radial extent of the $^{12}$CO(1--0) is about $60$~arcsec ($4.3$~kpc) in the NW, while it reaches to about $90$~arcsec ($6.5$~kpc) in the SE (see Fig.~\ref{fig:moments}). However, $^{12}$CO(3--2) is more centrally concentrated. It extends to about $40$~arcsec ($2.8$~kpc) on either side. The velocity field is mostly regular up to $\approx50$~arcsec from the centre, and then the gas seems to become more patchy in the SE part of the disc compared to the NW part (see moment maps in Figure~\ref{fig:moments}). This asymmetry in the south of the galaxy was also seen in H\,{\small I} \citep{the97}. It is interesting to see this asymmetry also in CO, since CO is more centrally concentrated and more correlated with the star-forming gas component. CO is, therefore, considered to be less affected by any galaxy interactions compared to H\,{\small I} \citep{lav97}. Although this is not certain, the asymmetry could be related to interactions and/or past mergers with nearby galaxies, i.e. NGC~5002, NGC~5005 and NGC~5014 \citep{the97}.

The $^{12}$CO(1--0) and $^{12}$CO(3--2) PVDs indicate a stronger emission at the outer edges of the disc (see Figure~\ref{fig:pvds}). This edge brightening is simply due to the line of sight effect, i.e. the galaxy has a high inclination. Additionally, the $^{12}$CO(3--2) PVD shows an increase in brightness in the north, indicating more warmer gas there, compared to the southern part (see Figure~\ref{fig:pvds}). As seen from Figure~\ref{fig:pvds}, at about $10$ arcsec from the centre, PVDs of both lines reach the flat part of the rotation curve at $\approx200$~km~s$^{-1}$, i.e. at the maximum rotational velocity of the galaxy. A total dynamical mass of $2.5\times10^{11}$~M$_{\odot}$ was estimated using H\,{\small I} data out to a radius of $420$~arcsec \citep{lav97}. As seen from Figure~\ref{fig:pvds}, the CO gas reaches the circular velocity of $200$~km~s$^{-1}$ (it is $217$~km~s$^{-1}$ after correcting for the inclination, see Table~\ref{tbl:ngc5033}) at about $10$~arcsec from the galaxy's centre. Using the CO circular velocity, we calculated a dynamical mass of $8.2\times10^{9}$~M$_{\odot}$ within the central region of size $1.4$~kpc (equivalently $20$~arcsec in diameter). 

When barred galaxies are viewed close to edge-on orientation the central bar dynamics are usually associated with characteristic X-shaped PVDs, i.e. two kinematic components \citep{sell93,ku95,bu99,mer99,topal16}. As seen from Figure~\ref{fig:pvds}, the PVDs of both lines reveal a rapidly rising velocity component followed by a flat rotation curve. This indicates no evidence of a large-scale bar in the centre of NGC~5033, but we cannot exclude that the galaxy might host a small central bar \citep{the97,kohno03}. While, because of large central line widths, integrated intensities peak at the centre (Figure~\ref{fig:moments}) when observed with $6$ or $14.5$~arcsec beams, the PVDs (Figure~\ref{fig:pvds}) reveal that the CO gas in the very centre of the galaxy shows a local minimum. This may be caused by the AGN (the $1.4$~GHz radio continuum flux density near the centre is $126$~mJy \citealt{the97}) or it is simply a result of lines-of-sight not crossing tangentially a spiral arm with comparatively strong molecular emission. Although the galaxy hosts an AGN, \citet{per07} concludes that the radio emission from the nuclear region of the galaxy is mainly powered by starburst activity (see also \citealt{yun01}). 
%

% Figure: Line Ratios comparison
%
\begin{figure*}
\centering
  \includegraphics[width=8cm,clip=]{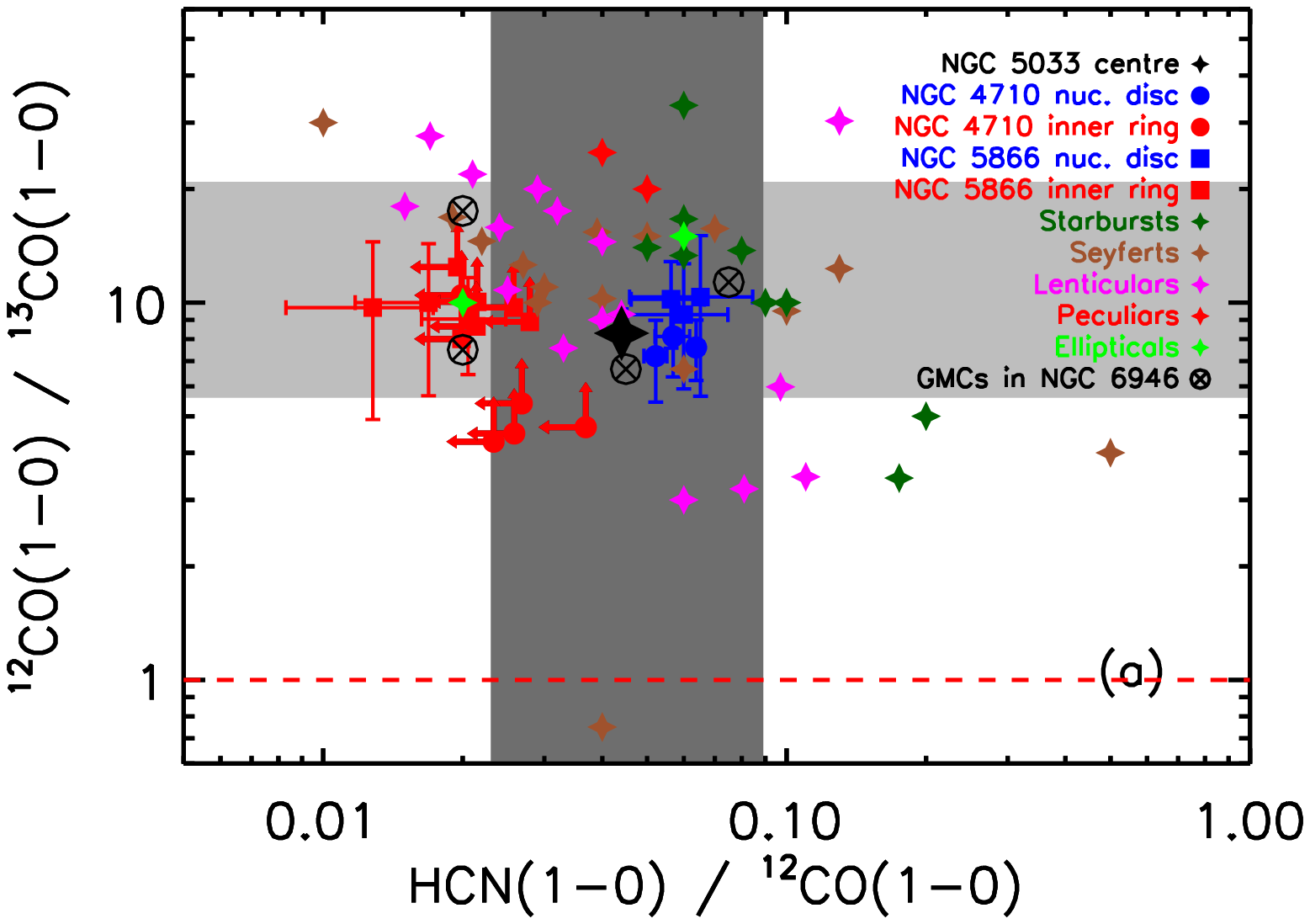}
  \includegraphics[width=8cm,clip=]{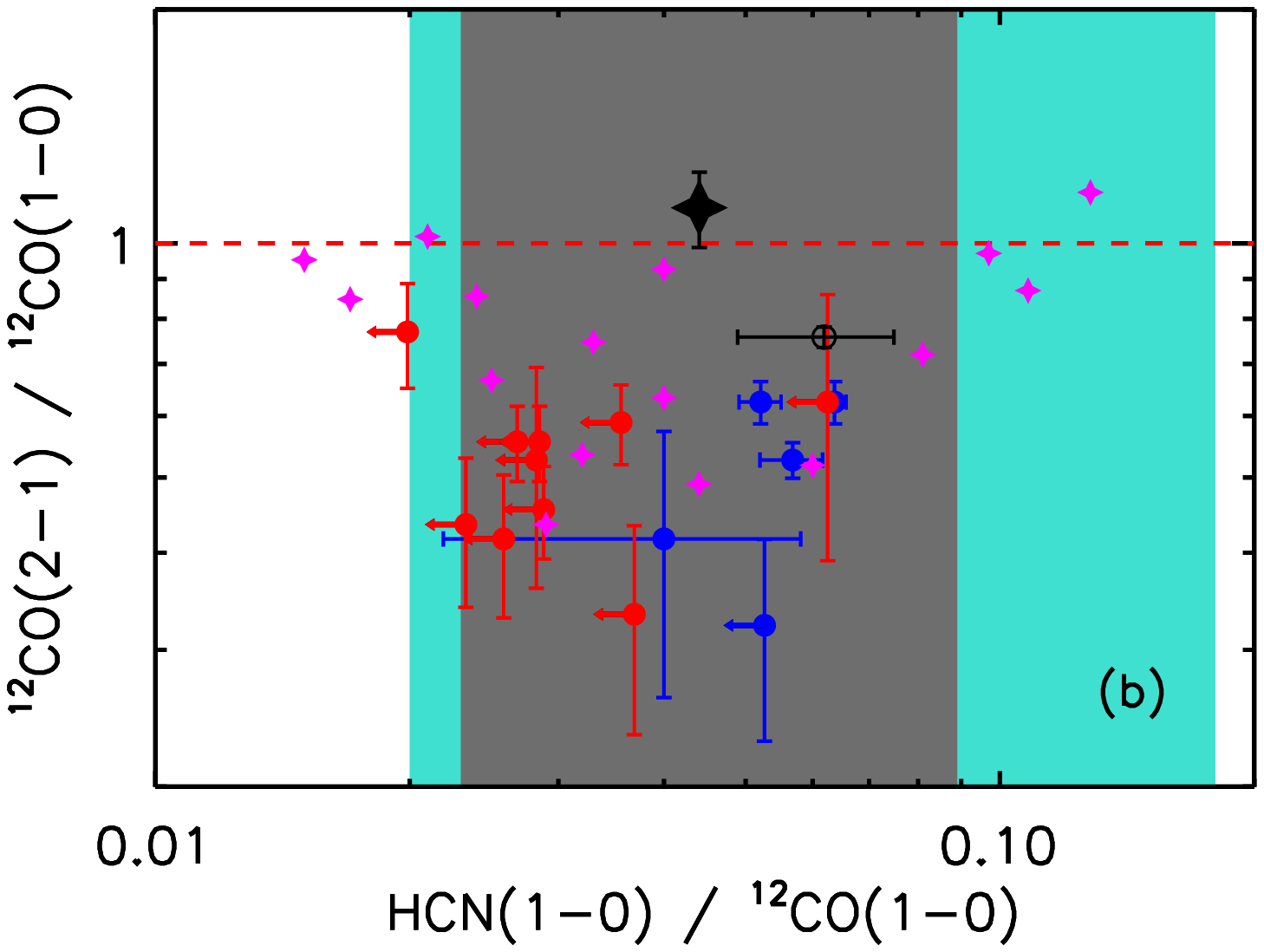} \\
  \includegraphics[width=8cm,clip=]{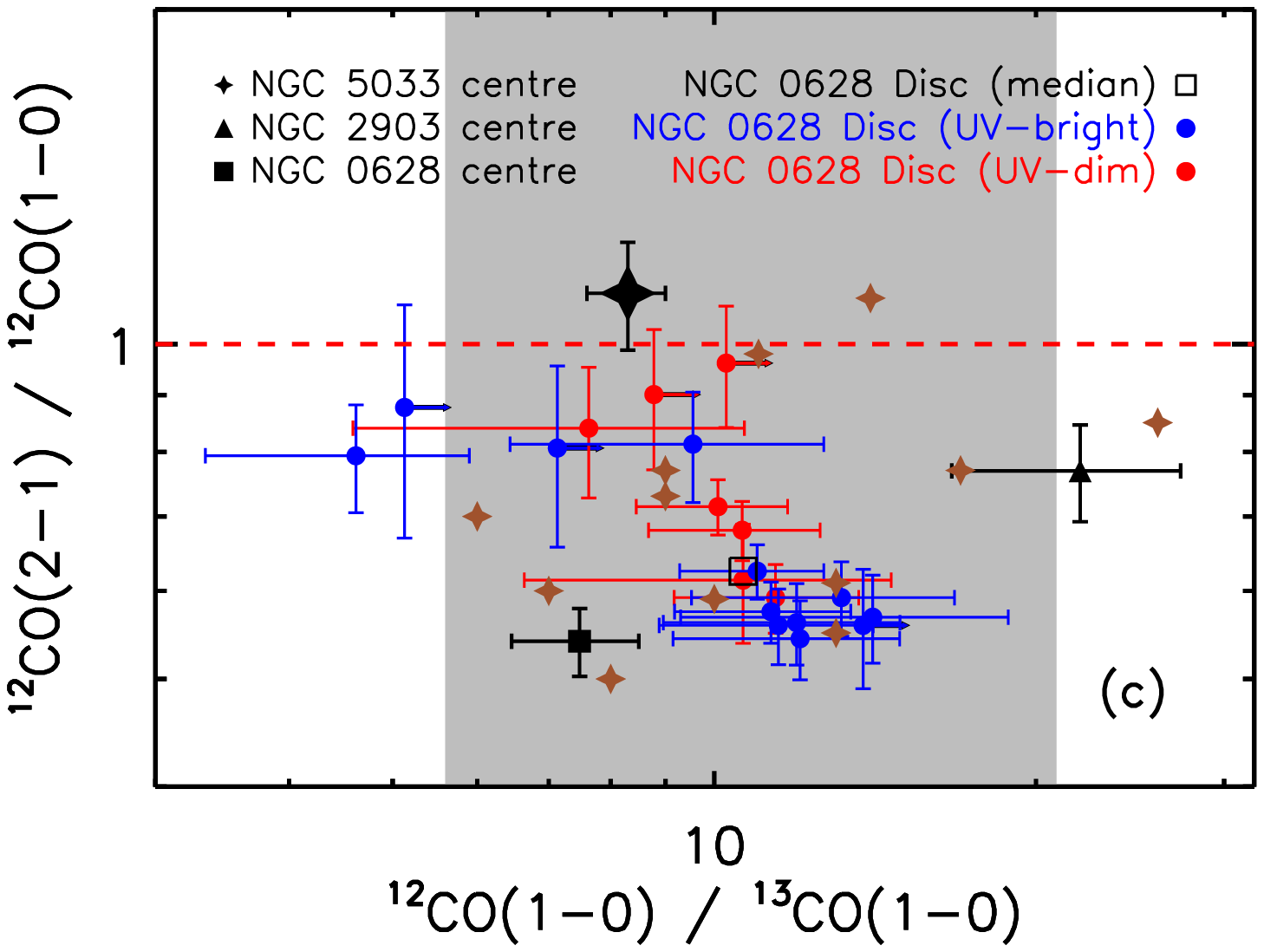} 
  \includegraphics[width=8cm,clip=]{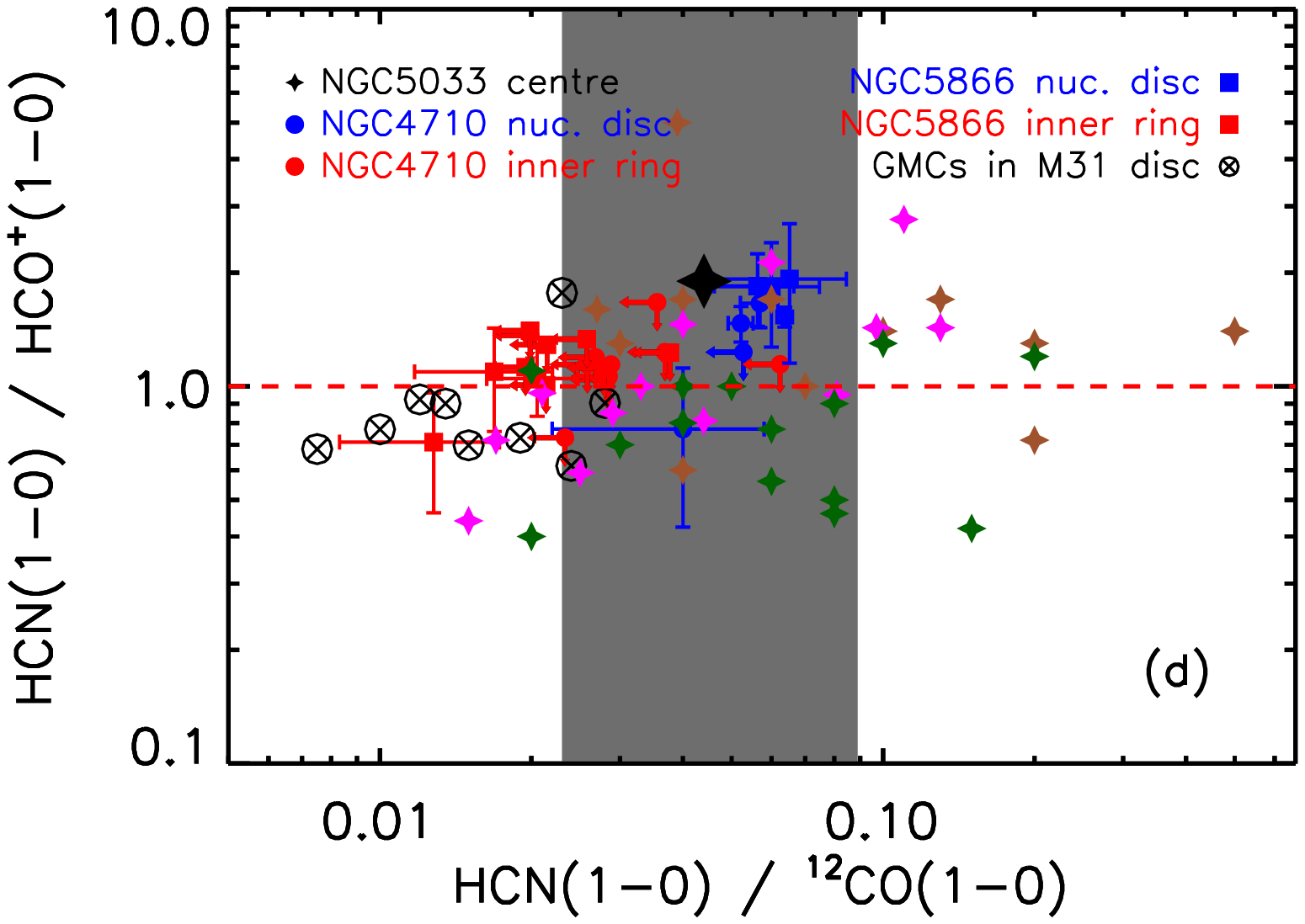} \\
  \caption{Black filled stars represent the ratios of the integrated line intensities (Table~\ref{tbl:ratiomod}) at the centre of NGC~5033 in all panels. Red and blue squares and circles represent the ratios in the inner ring and nuclear disc of NGC~5866 and NGC~4710 \citep{topal16}, respectively, in all panels, except panel $c$ (see the legend in the panel). The data points with arrows indicate lower and upper limits estimated following the same approach we performed in the current study (see Section~\ref{sec:inten}). Starburst, Seyfert, lenticular, peculiar and elliptical galaxies are shown by dark green, brown, magenta, red and green filled stars, respectively \citep{kri10}. The data for Seyfert galaxies in panel $c$ were taken from \citet{pa98}. Black circles with a cross in panels $a$ and $d$ represent the GMCs in NGC~6946 \citep{topal14} and in M31 \citep{bro05}, respectively. The light grey shaded areas in panels $a$ and $c$ show the range for the $R_{\rm 11}$ ratios in spirals \citep{pag01}, while the dark grey shaded area in panels $a$, $b$ and $d$ show the range for $R_{\rm DC}$ ratios in spirals \citep{gao04}. The turquoise shaded area in panel $b$ represents the ratios for starbursts \citep{ba08}. The open black square and circle in panel $b$ show the ratios in the centre of NGC~5866 and NGC~4710, respectively \citep{cr12}. The ratio of $1$ in all panels is indicated by red dashed line.}
  \label{fig:linerat}
\end{figure*}

\subsection{Line ratios}
\label{sec:ratio}
A $22$~arcsec ($1.6$~kpc) beam typically contains many GMCs, so any physical properties reflected by the line ratios will eventually represent average conditions for such complexes. We assume that the source is resolved (i.e. fills the beam) and the GMCs within the beam have the same average physical conditions. Since higher excitation of a molecule requires a medium with higher density and/or temperature, $^{12}$CO(3--2) indicates the existence of slightly denser and warmer gas compared to $^{12}$CO(1--0) and $^{12}$CO(2--1). 

Compared to its parent molecule $^{12}$CO, $^{13}$CO is generally optically thinner and less abundant, i.e. allowing us to identify regions with higher column densities. An increase in the $R_{\rm 11}$ = $^{12}$CO(1--0) / $^{13}$CO(1--0) ratio, therefore, indicates thinner CO gas and vice-versa. To the contrary, $^{12}$CO(1--0), as it is more abundant, has a better chance for self-shielding and can therefore better survive in a hostile interstellar environment. However, there is evidence that a considerable fraction of CO emission could arise from relatively diffuse and lower column density ($<10^{22}$~cm$^{-2}$) molecular gas, and a significant fraction of molecular gas could be in a non-star-forming phase \citep{gold08,hei10,krum12,pe13}. The gas could be more diffuse due to supernova explosions, stellar winds, intense ultraviolet (UV) radiation and associated H\,{\small II} regions \citep[e.g.][]{fu09}. Such feedback processes could be different in the arms, inter-arm regions or in the centre of a galaxy. In galaxies, centre, arms and inter-arm regions could, therefore, have different $R_{\rm 11}$ ratios, causing fluctuations in the $R_{\rm 11}$ ratio, generally seen across the disc of a galaxy \citep[e.g.][]{topal20}. 

Ratios between HCN to CO, i.e. $R_{\rm DC}$ = HCN(1--0) / $^{12}$CO(1--0), indicate the portion of dense gas in the medium (i.e. the higher the $R_{\rm DC}$ ratios, the higher the dense gas fraction), while $R_{\rm DD}$ = HCN(1--0) / HCO$^{+}$(1--0), provides information on the dominant physical processes in the environment. Supernova explosions, strong UV radiation from young massive stars, and X-rays from AGNs, all play an important role in defining physical properties of the ISM, and affect the molecular line ratios observed. 

Although UV radiation affects the outer layers of clouds and create photo-dissociation regions or PDRs \citep[e.g.][]{hol99}, X-rays can penetrate deeper into the clouds and create X-ray dissociation regions or XDRs \citep[e.g.][]{mal96}. Based on their theoretical approach \citet{mei07} found that when the density is $>$ $10^{5}$~cm$^{-3}$ $R_{\rm DD}$ $< 1$ for XDRs and $R_{\rm DD}$ $> 1$ for PDRs. However at lower densities $R_{\rm DD}$ gets higher than unity in XDRs. In galaxies with a high rate of supernova explosions, HCO$^{+}$, a molecule very sensitive to the ionization level of molecular gas, could be enhanced with respect to HCN (i.e. $R_{\rm DD} < 1$), due to cosmic rays (CRs) produced by young supernova remnants \citep{dick80,wot81,el83, jo98, cos11} Please note that in the central region of active galaxies the shock from the AGN could also produce CRs \citep{rac93, fer98, ach01, ber08, guo11}. The CRs in the medium could ionize the H$_{2}$ and produce H$_{3}$$^{+}$ which could react with CO to create HCO$^{+}$ \citep{pad09}. In light of this information, we will discuss the results of our line ratio diagnostics.\\

\subsubsection{Ratios between CO lines}
\label{sec:coratio}
As listed in Table~\ref{tbl:ratiomod}, the line ratios indicate that $^{12}$CO(2--1) is brighter than $^{12}$CO(1--0) and $^{12}$CO(3--2) at the centre of the galaxy. We compare the ratios found at the centre of NGC~5033 to that of different types of galaxies in Figure~\ref{fig:linerat}. A CO survey of about $80$ nearby spiral galaxies indicates an average ratio of $R_{\rm 21} = 0.89\pm0.06$ \citep{bra93}. Similarly, the range for $R_{\rm 21}$ ratios in nearby galaxies (including our own Milky Way) is from $0.6$ to $1$ \citep{hase97, ler09, topal20}, lower than the ratio of $1.1\pm0.1$ we find at the centre of NGC~5033. It is also seen in panels $b$ and $c$ of Figure~\ref{fig:linerat} is that the $R_{\rm 21}$ ratio at the centre of NGC~5033 is higher than in most of the galaxies, indicating a relatively warm and/or dense gas component. The panel $c$ of the same figure also indicates that NGC~5033 has a higher $R_{\rm 21}$ compared to the centre of the spiral galaxies NGC~0628 \citep{topal20} and NGC~2903 \citep{topal18} showing no AGN activity, again indicating relatively highly excited gas in the AGN host centre of the spiral galaxy NGC~5033. 

The $R_{\rm 31}$ ratios were obtained in the centre of $61$ galaxies from different Hubble types \citep{mao10}. They found that the $R_{\rm 31}$ ratio ranges from $0.2$ to $2$. While the $R_{\rm 31}$ ratio in starbursts could be as high as $2$, the central regions of normal spirals mostly have ratios at the lower end of this range \citep{mao10}. We find $R_{\rm 31} = 0.4\,\pm\,0.1$ at the centre of NGC~5033, which is within the range for spirals, but higher than the ratio in the centre of NGC~0628 (i.e. $R_{\rm 31} = 0.19\,\pm\,0.03$; \citealt{topal20}) and lower than the ratio in the centre of NGC~2903 (i.e. $R_{\rm 31} = 0.83\,\pm\,0.14$; \citealt{topal18}). This indicates that the active centre of the spiral galaxy should have warmer and/or denser (at least moderately) gas compared to most normal spirals, but more certainly colder than the centre of starbursts. It is seen from Figure~\ref{fig:linerat} that NGC~5033 has a relatively low $R_{\rm 11}$ ratio compared to most of the galaxies, including the range for normal spirals (see panels $a$ and $c$), most positions over the disc of lenticular galaxies NGC~4710 and NGC~5866 (see panel $a$), and those located along the disc of the spiral galaxy NGC~0628 (see panel $c$). This indicates relatively less tenuous central gas in NGC~5033.

\begin{figure*}
\centering
  \includegraphics[width=8cm,clip=]{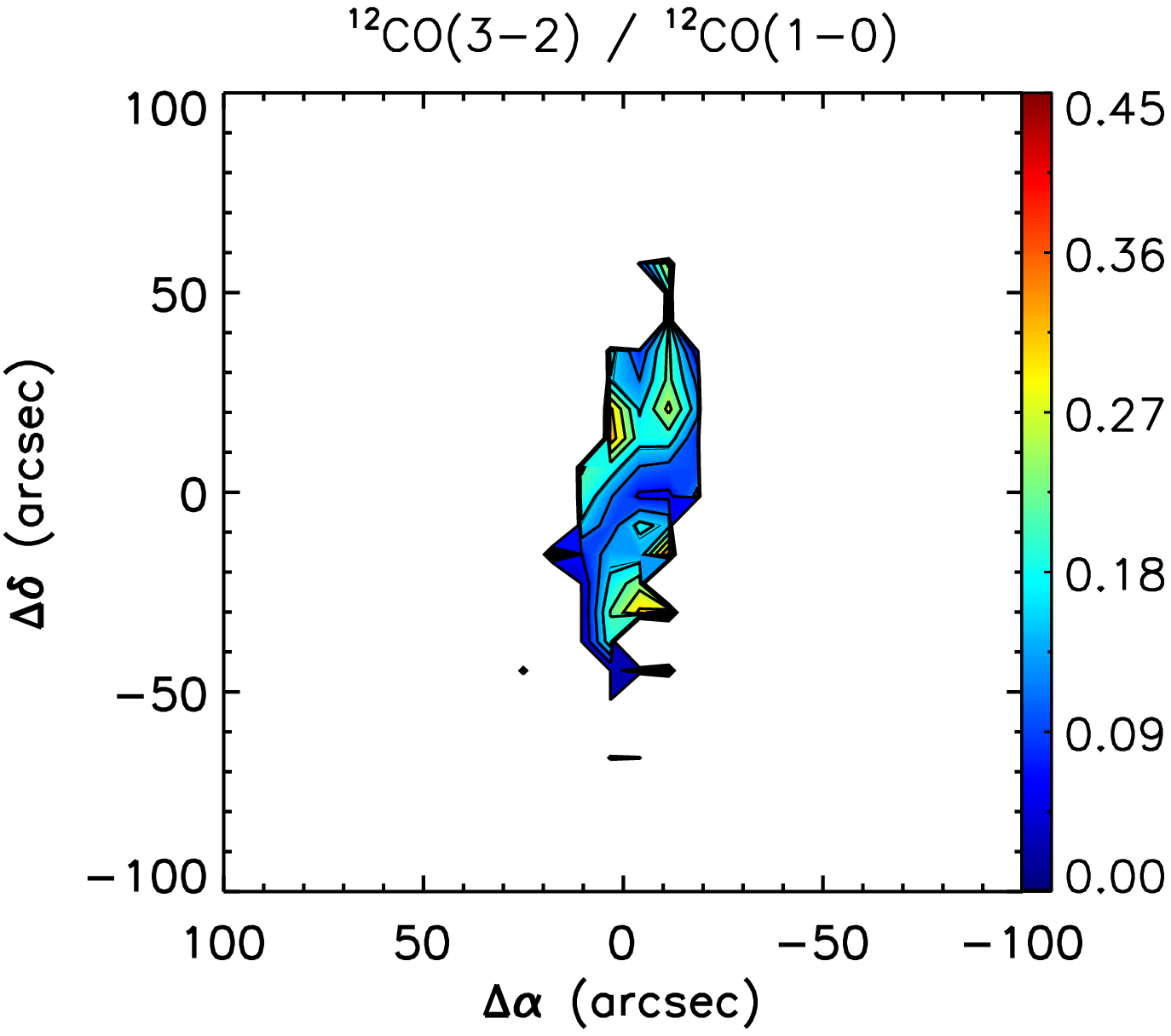}
  \includegraphics[width=8cm,clip=]{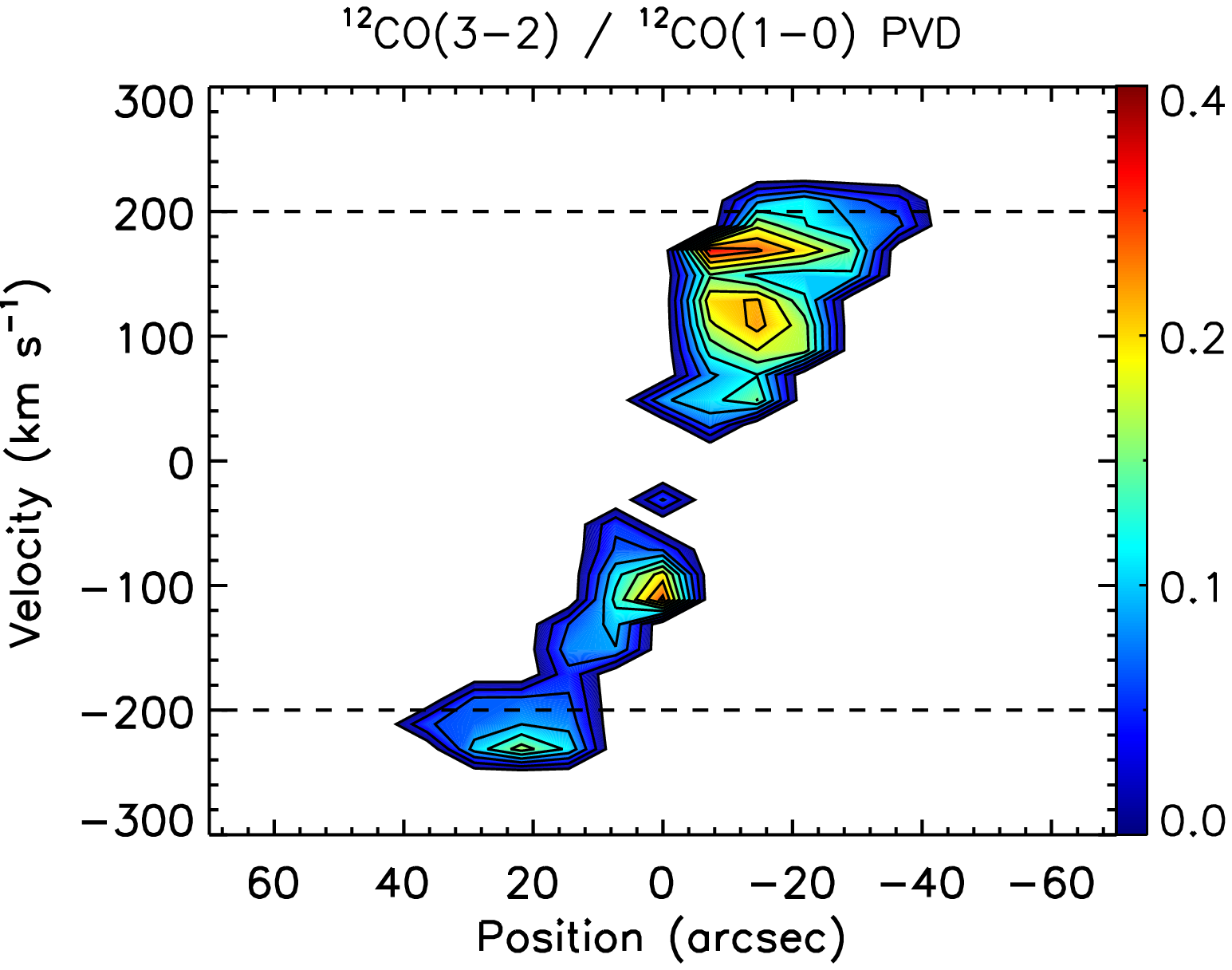} \\
  \caption{{\bf Left}: Contour map for the ratios of the $^{12}$CO(3--2) moment 0 map to that of $^{12}$CO(1--0). {\bf Right}: Contour map for the ratio of $^{12}$CO(3--2) PVD to that of the $^{12}$CO(1--0) PVD. On both ratio maps the contour levels start from $10$ per cent of the peak value with an increment of $10$ per cent. East to the left and north is up in both panels.}
  \label{fig:mompvd}
\end{figure*}

The $R_{\rm 31}$ ratio moment map and ratio PVD are shown in Figure~\ref{fig:mompvd}. As seen from the ratio moment map, the $R_{\rm 31}$ ratio is mostly higher in the north of the galaxy. This could be explained by the fact that $^{12}$CO(3--2) emission is generally fainter in the south (see Figures~\ref{fig:moments} and \ref{fig:pvds}) causing lower $R_{\rm 31}$ ratios (relatively colder and/or less dense gas) there compared to the north where the $R_{\rm 31}$ ratio is higher (see Figure~\ref{fig:mompvd}). The same difference is more clearly visible in the ratio PVD maps. i.e. the ratio is higher in the north of the galaxy (see Figure~\ref{fig:mompvd}). To summarize: both the moment map and the PVD indicate that the two sides of the disc have different average excitation conditions.

% Figure: Line Ratios
%

\subsubsection{Ratios between dense gas tracers}
\label{sec:denratio}
As seen from panel $d$ of Figure~\ref{fig:linerat}, $R_{\rm DD}$ $> 1$ in the centre of NGC~5033, while almost all starbursts (shown by green filled stars) and GMCs in M31 (shown by black circles with a cross) have $R_{\rm DD} < 1$.  This indicates that in the centre of NGC~5033, HCN could be enhanced and/or HCO$^{+}$ could be lower. \citet{krip08} studied the centres of $12$ active galaxies based on HCN and HCO$^{+}$ observations and found an average ratio of $R_{\rm DD}$ = $1.1\,\pm\,0.1$ (with a median of $1.2\,\pm\,0.1$) for their sample, which is almost half the value found at the centre of NGC 5033, i.e. $R_{\rm DD}$ = $1.9$ \citep{kri10}. However, \citet{pri15} found a mean ratio of $R_{\rm DD}$ = $1.84\,\pm\,0.43$ for a sample of AGN host galaxies, which is the same as the ratio found in the centre of NGC~5033.

As our model results suggest a density range of $n$(H$_{2}$)$ =10^{3} - 10^{4}$~cm$^{-3}$ (see Table~\ref{tbl:ratiomod} and Section~\ref{sec:modres}) and XDRs could have $R_{\rm DD}$ $> 1$ in low-density environments (see the discussion at the beginning of Section~\ref{sec:ratio}), this indicates that the existence of XDRs rather than PDRs is a more likely scenario for the line ratio observed in the centre of NGC~5033. Therefore, our high $R_{\rm DD}$ ratio could be explained by the enhancement of HCN in comparison with HCO$^{+}$ through X-rays near the AGN of NGC~5033 as this is the case in many AGN host Seyfert galaxies \citep[e.g.][]{koh03,ima07,dav12, pri15}. 

Other supportive evidences for the high $R_{\rm DD}$ ratio observed in the centre of NGC~5033 are the following. No supernovae have yet been found at the centre of the galaxy \citep{catalog17}. This supports the idea that any effects of supernova explosions on the central gas complex should not dominate potential AGN activity. Besides, the CRs created by the AGN in the centre of NGC~5033, which could cause a considerable decrease in the $R_{\rm DD}$ ratio, appear not to dominant over effects from X-ray radiation. The higher $R_{\rm DD}$ ratio observed at the centre is, therefore, possibly due to a lack of CRs, which might otherwise cause a higher abundance of HCO$^{+}$ in the medium. However, please note that, although no supernovae have yet been detected in the centre of the galaxy and we somehow do not see the effects of supernova in the line ratios, it is still possible for supernovae to be invisible due to high optical depths. 

Overall, the most likely scenario for the relatively high $R_{\rm DD}$ ratio found at the centre of NGC~5033 is the existence of an AGN. Effects of X-ray radiation may dominate in the medium, while CR and PDR chemistry may be less important in the heart of the galaxy.

\subsubsection{Ratios between CO to dense gas tracers}
\label{sec:codenratio}
As seen from panels $a$, $b$ and $d$ of Figure~\ref{fig:linerat}, the active centre of NGC~5033 has an average $R_{\rm DC}$ ratio compared to other types of galaxies, indicating a similar dense gas fraction. From the same panels, the $R_{\rm DC}$ ratio found at the centre of NGC~5033 is similar to that seen in the nuclear disc but higher than in the inner rings of the lenticular galaxies NGC~4710 and NGC~5866 \citep{topal16}. The ratio of $R_{\rm DC}$ = $0.044\,\pm\,0.001$ found in the centre of NGC~5033 is within the range found in the centre of the Seyfert galaxy NGC~3227 ($R_{\rm DC}$ = $0.1$ to $0.01$). Given the AGN host nature of NGC~5033, it is plausible to assume that high X-ray ionization rates in the central region could increase the HCN abundance \citep[e.g.][]{use04,mei05}, causing the $R_{\rm DC}$ ratio to increase. Intensity ratios of $^{12}$CO/HCO$^{+}$$\,\approx\,43$ in the centre of NGC~5033 \citep{kri10} are similar to the ratios found in other spiral galaxies, i.e. $^{12}$CO/HCO$^{+}$$\,>\,25$ \citep{jim19}.This indicates that the effects of supernova explosions could not be dominant in the centre of NGC~5033, as already suggested by the $R_{\rm DC}$ ratio (see Section~\ref{sec:denratio}).

\subsection{Modelling}
\label{sec:modres}

The ISM has very complex characteristics with different temperature, density, radiation field, and mechanical feedback. It is, therefore, only natural to assume the existence of multiple molecular components in the gaseous ISM. In such a complex structure, a combination of low-$J$ CO lines with a $1$D model (such as {\sc RADEX}) can only provide us with some information on relatively colder and tenuous gas in the ISM. Transitions between higher energy levels of CO (i.e. $J~\ge~4-3$) are necessary to probe the physical properties of a hotter gas component with a broad range of densities. Using a complete set of CO lines along with multiple dense gas tracers could, therefore, enable us to model multiple gas components of the ISM and provide a better insight on multiple phases of star-forming gas clouds \citep[e.g.][]{topal16}. Nevertheless, given the number of CO lines available to us (i.e. $^{12}$CO(1--0, 2--1, 3--2) and $^{13}$CO(1--0, 2--1)), this study still provides us with valuable information on the physics of relatively colder and tenuous gas in the centre of NGC~5033. In consideration of this information, we discuss our model results as follows.

\begin{figure}
\centering
  \includegraphics[width=7cm,clip=]{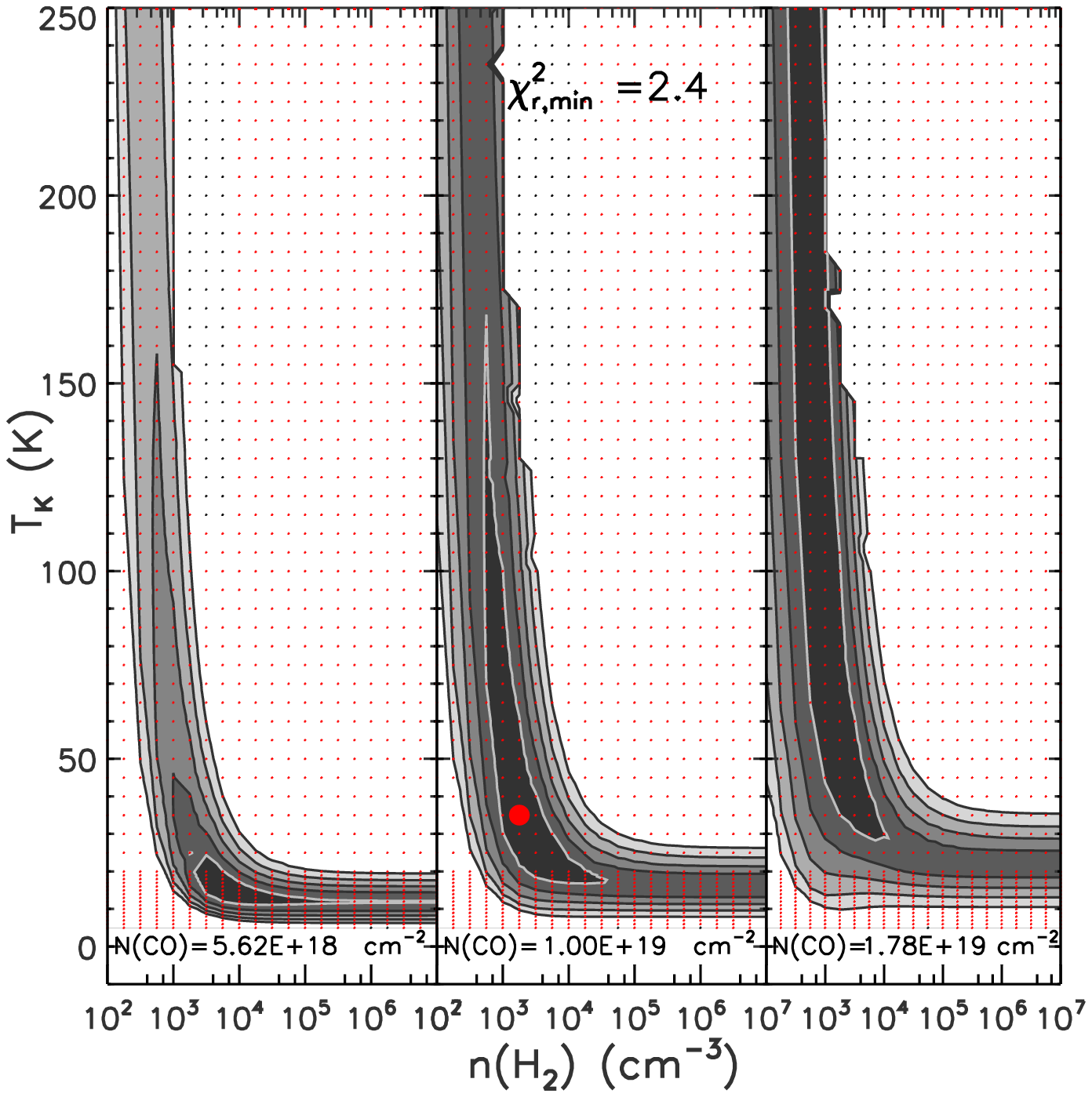}
  \includegraphics[width=6.7cm,clip=]{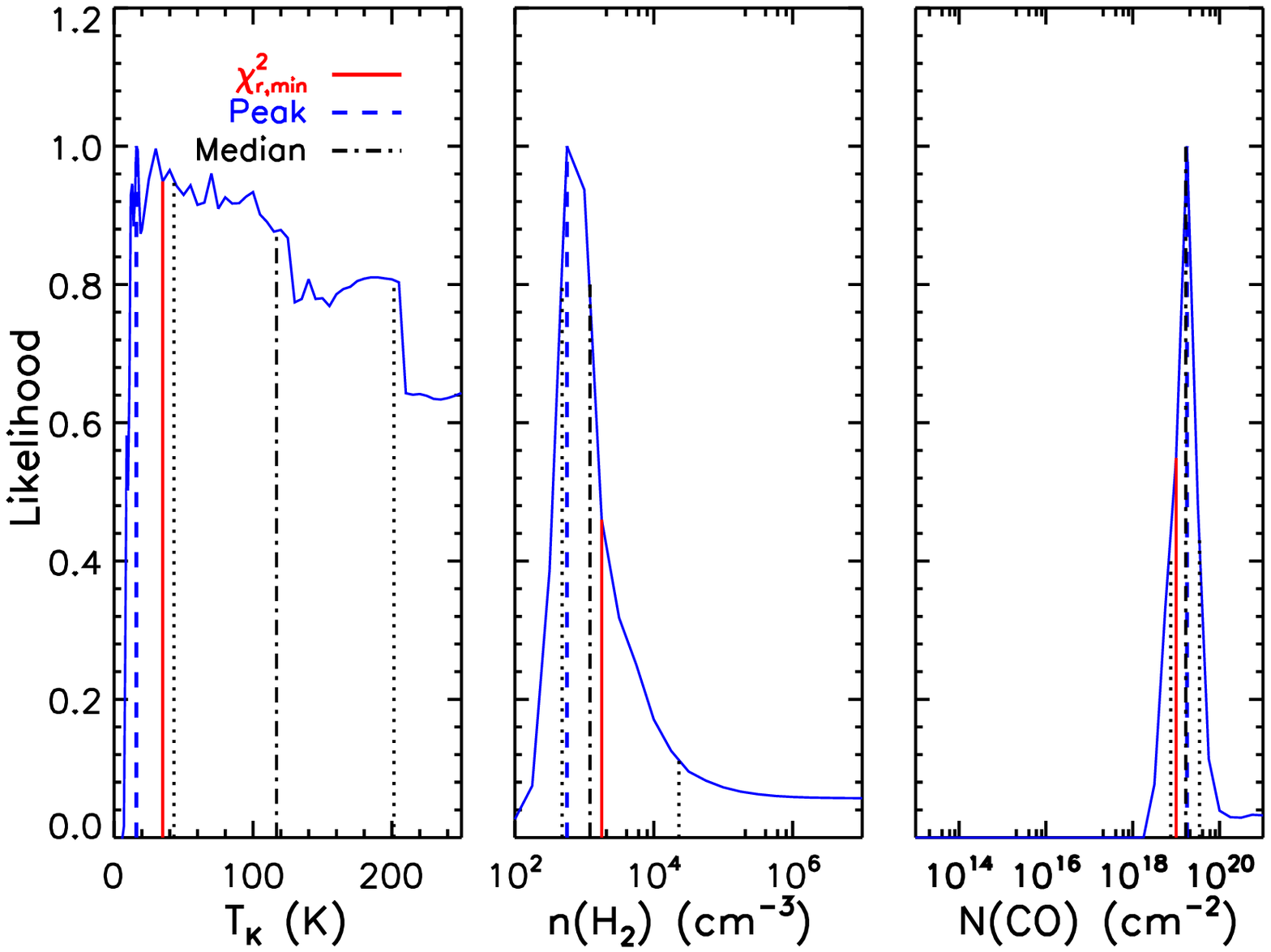}\\
  \caption{{\bf Top:} $\Delta\chi_{\rm r}^{2}$ = $\chi_{\rm r}^2$ - $\chi_{\rm r, min}^2$ maps are shown. 
  $\Delta\chi_{\rm r}^{2}$ is given as a function of $T_{\rm K}$ and $n$(H$_2$), for three values of $N$(CO) centred around the best fit. The values of $N$(CO) are shown at the bottom of each panel, while the value of reduced $\chi_{\rm min}^2$, i.e. $\chi_{\rm r, min}^2$, of the best fitting model is given in the central panel and is indicated by a red filled circle. Small red dots indicate values considered in the model parameter space, whereas small black dots represent unused bad models (see \citealt{van07}). The contour confidence levels for $3$ degrees of freedom are $3.5$, $8.0$, $14.2$, $22.1$ and $31.8$ for $1\sigma$ ($68\%$ probability; the darkest greyscale), $2\sigma$, $3\sigma$, $4\sigma$ and $5\sigma$, respectively. The $1\sigma$ confidence levels are indicated by grey lines, while the levels higher than $1\sigma$ are shown by black lines. {\bf Bottom:} Probability distribution function (PDF) of each model parameter marginalized over the other two (see Section~\ref{sec:bestlvg}). The peak (most likely) is indicated with a dashed blue line, whereas the median and the $1\sigma$ confidence around it are shown with a dash-dotted blue and dotted black line, respectively. The solid red line indicates the best-fitting model in a $\chi^{2}$ sense.}
  \label{fig:chilike}
\end{figure}

As seen from Figure~\ref{fig:chilike}, the $\Delta\chi_{\rm r}^{2}$ contour maps indicate that $T_{\rm K}$ increases sharply at lower $n$(H$_{2}$), and it gradually decreases as $n$(H$_{2}$) increases (i.e. the banana-shaped degeneracy, see \citealt{van07,topal14, topal20}). Single Gaussian profiles are seen in the PDFs (see Section~\ref{sec:bestlvg}), except that for $T_{\rm K}$ (Fig.~\ref{fig:chilike}, lower left panel), causing the best-fitting and most likely model results being in agreement, i.e. the best-fitting model is contained within the $68$ per cent ($1\sigma$) confidence levels around the median for $n$(H$_{2}$) and $N$(CO).

%
% Figure: chi-square contour map
%

According to our best model results (in a $\chi^{2}$ sense), NGC~5033 is hosting a central molecular gas complex with physical properties as follows: $T_{\rm K}~=~35$~K, $\log$($n$(H$_2$))$~=~3.3$~cm$^{-3}$, and $\log$($N$(CO))$~=~19.0$~cm$^{-2}$ (see Table~\ref{tbl:ratiomod}). The most likely results in agreement with the $\chi^{2}$ results except for $T_{\rm K}$ (i.e. the most likely result indicates a higher temperature compared to the temperature indicated by the $\chi^{2}$ method) are also listed in Table~\ref{tbl:ratiomod}. We compare our model results to those in the literature performed with the same radiative transfer code and using low-$J$ CO lines similar to our study. 

Many studies have probed the physics of the gas in the flocculent spiral galaxy NGC~6946 \citep{ib01,wal02,ba06,topal14}. \citet{ib01} found two distinct molecular components in the centre of NGC~6946, i.e. a warm and dense, and a hotter and more tenuous one. Two other studies found similar gas components, i.e. a warm ($T_{\rm K}~=~40$~K) and a hotter one ($T_{\rm K}~=~130$~K), with similar hydrogen volume number densities, i.e. $\log$($n$(H$_2$))~$\approx3.3$~cm$^{-3}$ \citep{wal02, ba06}. According to our model results, the central gas complex in NGC~5033 has the same physical properties as the central warm gas component in NGC~6946 with similar overall temperature, density and column density.

%
%Model results
%		
 \begin{table*}
\begin{center}
 \small
 \caption{Line ratios and model results from the central $22$~arcsec area.} %% no full stop at the end of caption
 \begin{tabular}{cclll}
\hline  %% rule at top
            Line ratio$^{a}$& Value & Parameter & $\chi^{2}$ & Likelihood \\ 
            \hline 
  $R_{\rm 21}$&$\phantom{0}1.1\pm0.1$&$T_{\rm K}$&$35$~K&$117^{+84}_{-74}$~K\\
  $R_{\rm 31}$&$\phantom{0}0.4\pm0.1$&$\log$($n$(H$_{\rm 2}$))&$3.3$~cm$^{-3}$&$3.1^{+1.3}_{-0.4}$~cm$^{-3}$\\
  $R_{\rm 11}$&$\phantom{0}8.3\pm0.7$&$\log$($N$(CO))&$19.0$~cm$^{-2}$&$19.2^{+0.3}_{-0.3}$~cm$^{-3}$\\
  $R_{\rm 22}$&$\phantom{0}7.2\pm1.1$&$[^{12}\rm C]/[^{13}C]$&$20$&\\
  $R_{\rm DC}$&$\phantom{0}0.044\pm0.001$&&&\\
  $R_{\rm DD}$&$1.9$&&&\\  	
 \hline %% rule at bottom
 \end{tabular}
 \label{tbl:ratiomod}
 \end{center}
 $^{\rm a}$ Integrated line intensity (K~km~s$^{-1}$) ratios of $R_{\rm 21}$, $R_{\rm 31}$, $R_{\rm 11}$, $R_{\rm DC}$, and $R_{\rm DD}$ stand for the ratios of $^{12}$CO(2--1) / $^{12}$CO(1--0), $^{12}$CO(3--2) / $^{12}$CO(1--0), $^{12}$CO(1--0) / $^{13}$CO(1--0), HCN(1--0) / $^{12}$CO(1--0), and HCN(1--0) / HCO$^{+}$(1--0), respectively.
 \end{table*}
 
\citet{topal16} probed physical properties of molecular gas in the lenticular galaxies NGC~4710 and NGC~5866 using low-$J$ CO lines, and the same radiative transfer code. Our model results indicate that there is warmer gas with similar $n$(H$_2$) and $N$(CO) in the centre of NGC~5033 compared to the centre of NGC~4710. The model results agree with the empirical results (i.e. the results inferred directly from the observed line ratios without modelling, see Section~\ref{sec:ratio}) in terms of temperature. However, the gas in the centre of NGC~5866 has quite different properties compared to the centre of NGC~5033, i.e. the gas is warmer and denser with higher column densities compared to the central region of NGC~5033. This indicates a disagreement between the empirical (see Section~\ref{sec:ratio}) and model results. 

The physics of the star forming gas at the heart of the spiral galaxy NGC~2903 was also studied by using low-$J$ CO lines and {\sc RADEX} \citep{topal18}. NGC~2903 hosts denser and colder gas in its centre (i.e. $T_{\rm K}~=~20$~K and $\log$($n$(H$_2$))$~=~4.2$~cm$^{-3}$) with the same $N$(CO) compared to the centre of the Seyfert galaxy NGC~5033. Additionally, various positions over the disc of the spiral galaxy NGC~0628 were studied using an identical method to ours \citep{topal20}. They found that the central gas complex in NGC~0628 has the following physical properties; $T_{\rm K}~=~30$~K, $\log$($n$(H$_2$))$~=~2.2$~cm$^{-3}$, and $\log$($N$(CO))$~=~17.7$~cm$^{-2}$. These results indicate that the gas is warmer and denser with higher $N$(CO) in the centre of NGC~5033 compared to the centre of NGC~0628. Our model and empirical results (see Section~\ref{sec:ratio}) are in agreement in the sense that the centre of NGC~5033 hosts warmer gas than the centres of NGC~2903, NGC~0628, and NGC~4710. 

\begin{figure}
\begin{center}
  \includegraphics[width=9cm,clip=]{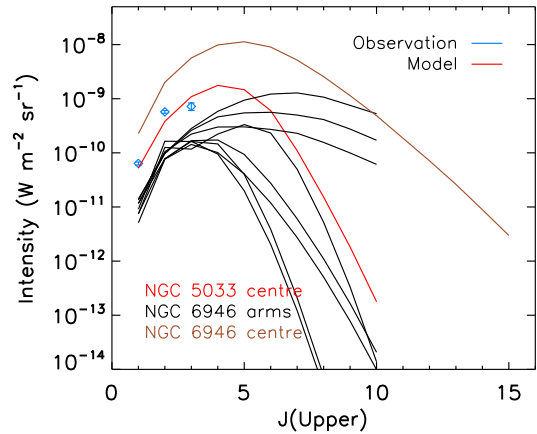}
  \caption{CO SLEDs. 
  The total intensity is shown as a function of the upper energy level of CO transitions up to $J$ = $15-14$. 
  The solid red line shows the best-fitting model, 
  while the blue diamonds with error bars represent the observations for NGC~5033. 
  SLEDs obtained in the centre \citep{ba06} and along the arms of NGC~6946 \citep{topal14} 
  are shown by solid brown and solid black lines, respectively, for comparison.}
  \label{fig:opt}
  \end{center}
\end{figure}

The turnover of the spectral line energy distributions, SLEDs, (hereafter $J_{\rm max}$) provides us with valuable information on SF activity. It is known that heavily star-forming high-$z$ or local galaxies could have $J_{\rm max}~>~6$, while  $J_{\rm max}~=~4$ in the centre of our own Milky Way \citep{fix99,ba04,we05,weis07}. As an extreme example, the heavily star-forming Orion Bar photodissociation region in our Galaxy has $J_{\rm max}~=~13$ \citep{hab10}. \citet{topal14} studied multiple regions over the patchy arms of NGC~6946 and obtained SLEDs for each region studied using the RADEX code. Similarly, \citet{ba06} obtained a SLED for the centre of NGC~6946 using the same LVG code. While \citet{topal14} found $J_{\rm max}~=~2-7$ (most regions have $J_{\rm max}~<~4$) across the disc of NGC~6946, \citet{ba06} found $J_{\rm max}~=~6$ at the centre of NGC~6946 (see Fig.~\ref{fig:opt}). 

As shown in Figure~\ref{fig:opt}, based on the best-fitting model $J_{\rm max}~=~4$ for the centre of NGC~5033, higher than that in most regions over the disc of NGC~6946, but lower compared to the centre of NGC~6946 where $J_{\rm max}~=~6$ (see Fig.~\ref{fig:opt}.) This indicates a higher (lower) activity at the heart of NGC~5033 compared to the spiral arms (the centre) of NGC~6946. However, the number of lines available to construct the SLED could also affect the location of the turnover. A CO ladder with more lines could, therefore, lead to a higher $J_{\rm max}$. The ISM has a complex multiphase structure, and divergence could be present in the SLED beyond $J~>~4-3$ \citep{papa14}. Any information from a single turnover representing the whole SLED will, therefore, provide average physical properties of such gas complexes. The turnover based on the low-$J$ transitions, consequently, offers physical properties averaged over relatively colder and tenuous gas components.

%
% Figure: optical depths
%

\section{Conclusion}
\label{sec:conc}

We probed the physics and kinematics of the molecular gas in the AGN host (Seyfert 1.8) spiral galaxy NGC~5033 using multiple low-$J$ CO lines [i.e. $^{12}$CO(1--0, 2--1, 3--2) and $^{13}$CO(1--0, 2--1)] and high-density tracers [i.e. HCN(1--0) and HCO$^{+}$(1--0)]. We also applied a radiative transfer code and obtained the best model results by comparing the modelled and the empirical line ratios. Our main conclusions are summarized below.

\begin{itemize}
\item $^{12}$CO(1--0) is more extended over the galaxy's disc compared to the other CO lines available to this study. At the centre of the galaxy $^{12}$CO(2--1) is the brightest among the CO transitions analysed by the current study. CO shows higher excitation (in terms of $3-2/1-0$ line ratios) northeast than southwest of the nuclear region. $^{12}$CO(1--0) gas extends further to the SE than to the NW from the centre, indicating an asymmetry in gas distribution. Previously published H\,{\small I} data confirm this asymmetry. Although determining the reason behind this asymmetry is not within the scope of the current study, we can still speculate that it could be related to past or ongoing interaction with nearby galaxies, i.e. NGC~5002, NGC~5005 and NGC~5014. 

\item The CO(1--0) integrated intensity shows a radial gradient, i.e. decreasing with the distance from the galactic centre. H$_2$ mass and gas surface density also decrease beyond $1$~kpc from the galactic centre. However, within the central $1$~kpc there is a decrease in H$_2$ mass and gas surface density because of the assumed depression in $X_{\rm CO}$, the conversion factor between H$_2$ column density and $^{12}$CO(1--0) integrated intensity. Position velocity diagrams (PVDs) reveal a central rapidly rising velocity component (possibly no large-scale bar). The integrated line intensity ratio of the moment 0 maps and the corresponding ratios of the PVDs indicate that the gas on either side of the disc has a different temperature, i.e. the gas is colder in the south. 

\item Line ratio diagnostics suggest that the centre of the spiral galaxy NGC~5033, hosting an AGN, harbours warmer and less tenuous gas with similarly dense gas fraction (i.e. similar $R_{\rm DC}$) with respect to the centres and the arms of most galaxies (including spirals, lenticulars and Seyferts). However, the gas in NGC~5033 is colder compared to starbursts. The $R_{\rm DD}$ = HCN(1--0) / HCO$^{+}$(1--0) $> 1$ ratio may be caused by the AGN, so the X-ray radiation in the medium could be the reason for this ratio, while CRs and PDRs seem to be less likely or at least not the primary driver of the ISM at the heart of the galaxy.

\item Our best model results (in a $\chi^{2}$ sense) mostly agree with the most likely model results, except for $T_{\rm K}$ since $T_{\rm K}$ results from the likelihood method are more uncertain (the likelihood method also indicates a higher temperature). The best model results for the centre of NGC~5033 are as follows: $T_{\rm K} = 35$~K, $\log$($n$(H$_2$)$) = 3.3$~cm$^{-3}$, and $\log$($N$(CO)$) = 19.0$~cm$^{-2}$. We compared our results to similar studies in the literature which used almost identical CO lines and the same radiative transfer code. The centre of the spiral galaxy NGC~5033 with its AGN hosts very similar gas to the warm gas component in the centre of the spiral galaxy NGC~6946. The gas is warmer in the centre of NGC~5033 compared to the centre of the spiral galaxies NGC~0628 and NGC~2903. However, the gas is denser with higher $N$(CO) compared to NGC~0628 and less dense with the same $N$(CO) compared to that in the centre of NGC~2903. Our model results suggest warmer gas with similar densities compared to the centre of the lenticular galaxy NGC~4710. However, the central gas in NGC~5033 is colder and/or less dense with lower $N$(CO) compared to the centre of the dusty lenticular galaxy NGC~5866. Our empirical and model results are in agreement in the sense that both methods suggest the existence of warmer gas in the centre of NGC~5033 compared to most spirals and lenticulars, including the spirals NGC~0628 and NGC~2903 and the lenticular galaxy NGC~4710. The SLED turnover, i.e. $J_{\rm max}$, for the centre of NGC~5033 occurs at a higher (lower) transition compared to most regions over the disc (the centre) of NGC~6946, indicating a relatively higher (lower) SF activity in the centre of NGC~5033 with respect to the disc (the centre) of NGC~6946.

\end{itemize}

 %% Acknowledgements
%
\section*{Acknowledgements}
ST would like to thank the anonymous referee for his/her insightful comments and suggestions. ST thanks Timothy A. Davis for insightful discussions on gas kinematics in galaxies. ST thanks Sugata Kaviraj for providing the code used to estimate the probability distribution function. The James Clerk Maxwell Telescope is operated by the East Asian Observatory on behalf of The National Astronomical Observatory of Japan; Academia Sinica Institute of Astronomy and Astrophysics; the Korea Astronomy and Space Science Institute; Center for Astronomical Mega-Science (as well as the National Key R\&D Program of China with No. 2017YFA0402700). Additional funding support is provided by the Science and Technology Facilities Council of the United Kingdom and participating universities in the United Kingdom and Canada. We acknowledge the usage of the HyperLEDA database (http://leda.univ-lyon1.fr). 

\section*{Data availability}
The data underlying this article will be shared on a reasonable request to the corresponding author.

\bibliographystyle{mn2e}
\bibliography{reference}
%
% Appendices
%
%\appendix

%
\end{document}